\documentclass[apj,tighten,iop,twocolumn]{emulateapj}
\usepackage{natbib}

\newcommand{\gtsimeq}{\raisebox{-0.6ex}{$\,\stackrel 
        {\raisebox{-.2ex}{$\textstyle >$}}{\sim}\,$}}

\newcommand{\lya}{Ly$\alpha$}

\newcommand{\oii}{[O\,{\sc ii}]}

\newcommand{\myemail}{chris.willott@nrc.ca}

\def\hst{{\it Hubble Space Telescope~}}
\def\spitzer{{\it Spitzer Space Telescope~}}
\def\co21{CO\,(2-1)}

\shorttitle{Exponential decline at the bright end of $z=6$ galaxy luminosity function}
\shortauthors{Willott et al.}

\begin{document}


\title{An exponential decline at the bright end of the $z=6$ galaxy luminosity function}


\author{
Chris J. Willott\altaffilmark{1},
Ross J. McLure\altaffilmark{2},
Pascale Hibon\altaffilmark{3},
Richard Bielby\altaffilmark{4},
Henry J. McCracken\altaffilmark{5},
Jean-Paul Kneib\altaffilmark{6},
Olivier Ilbert\altaffilmark{6},
David G. Bonfield\altaffilmark{7},
Victoria A. Bruce\altaffilmark{2},
and Matt J. Jarvis\altaffilmark{7},
}

\altaffiltext{1}{Herzberg Institute of Astrophysics, National Research Council, 5071 West Saanich Rd, Victoria, BC V9E 2E7, Canada; \myemail}
\altaffiltext{2}{SUPA Institute for Astronomy, University of Edinburgh, Royal Observatory, Edinburgh EH9 3HJ, UK}
\altaffiltext{3}{Gemini Observatory, Gemini South, AURA/Chile, P.O. Box 26732, Tucson, AZ 85726, USA}
\altaffiltext{4}{Dept. of Physics, Durham University, South Road, Durham, DH1 3LE, UK}
\altaffiltext{5}{Institut d'Astrophysique de Paris, UMR7095 CNRS, Universit\'e Pierre et Marie Curie, 98 bis Boulevard Arago, 75014 Paris, France}
\altaffiltext{6}{Laboratoire d'Astrophysique de Marseille, Universit\'e  Aix-Marseille,	38 rue Fr\'ed\'eric Joliot-Curie, 13388	Marseille, France}
\altaffiltext{7}{Centre for Astrophysics, Science and Technology Research Institute, University of Hertfordshire, Hatfield, Herts AL10 9AB, UK}

\begin{abstract}

We present the results of a search for the most luminous star-forming
galaxies at redshifts $z\approx 6$ based on CFHT Legacy Survey
data. We identify a sample of 40 Lyman break galaxies brighter than
magnitude $z'=25.3$ across an area of almost 4 square
degrees. Sensitive spectroscopic observations of seven galaxies
provide redshifts for four, of which only two have moderate to strong
\lya\ emission lines. All four have clear continuum breaks in their
spectra. Approximately half of the Lyman break galaxies are spatially
resolved in 0.7 arcsec seeing images, indicating larger sizes than
lower luminosity galaxies discovered with the {\it Hubble Space
  Telescope}, possibly due to on-going mergers. The stacked optical
and infrared photometry is consistent with a galaxy model with stellar
mass $\sim 10^{10}\, {\rm M}_\odot$.  There is strong evidence for
substantial dust reddening with a best-fit $A_V=0.75$ and $A_V>0.48$
at $2\sigma$ confidence, in contrast to the typical dust-free galaxies
of lower luminosity at this epoch. The spatial extent and spectral
energy distribution suggest that the most luminous $z\approx 6$
galaxies are undergoing merger-induced starbursts. The luminosity
function of $z=5.9$ star-forming galaxies is derived. This agrees well
with previous work and shows strong evidence for an exponential
decline at the bright end, indicating that the feedback processes
which govern the shape of the bright end are occurring effectively at
this epoch.

\end{abstract}


\section{Introduction}

The light from distant galaxies brings evidence of the conditions and
physical processes at play in the early Universe. By studying the
changes in galaxy properties over cosmic time we obtain a deeper
understanding of how our Universe evolved. At 1 billion years after
the Big Bang, galaxies were typically smaller, less luminous and less
dusty than today (Bouwens et al. 2006). Such observations need to be
explained by cosmological simulations which account for the
hierarchical merging of dark matter halos and the gas accretion and
cooling inside them necessary to form stars (e.g. Finlator et
al. 2011).

One of the key measurements of the evolving Universe is the galaxy
luminosity function. This function is related to the star formation
rate occurring at an epoch and how the star formation is distributed
across the galaxy population. Ultraviolet luminosity is well
correlated with the formation rate of young, hot stars, with the
caveat that dust extinction, common in starbursts, reduces the
observed ultraviolet flux. Comparison of the observed luminosity
function with that predicted by models constrains the important
physical processes occurring. For example, the faint end slope of the
galaxy luminosity function at redshifts up to at least $z=6$ is
flatter than the dark matter halo mass function (Bouwens et al. 2007),
which could be explained by feedback from supernovae winds (Cole 1991)
or from photoevaporation and heating during reionization (Barkana \&
Loeb 2001). At the bright end, the galaxy luminosity declines much
more sharply than the halo mass function. This is usually ascribed to
AGN feedback and inefficient gas cooling in high mass halos
(e.g. Benson et al. 2003).

Surveys for UV-continuum-selected galaxies at $z \gtsimeq 6$ have been
largely focused on \hst due to the small sizes of the galaxies and low
background in space. Hundreds of such galaxies have been identified in
deep ACS imaging of the GOODS, Hubble Ultra Deep (HUDF) and parallel
fields (Dickinson et al. 2004; Bunker et al. 2004; Yan \& Windhorst
2004; Bouwens et al. 2006).  The extremely deep imaging over the very
small area of the HUDF identifies galaxies with absolute magnitude at
1350\,\AA, $M_{1350}=-18$, corresponding to a star formation rate of
only $1\, $M$_\odot \,{\rm yr}^{-1}$ (Kennicut 1998). These \hst surveys have led to a
good description of the $z=6$ luminosity function over the range
$-21<M_{1350}<-18$ (Bouwens et al. 2007, 2008; Su et al. 2011). The
luminosity function can be fit by a Schechter function with
characteristic break luminosity of $M_{1350}=-20.2$. However, the few sources
detected brighter than the break leave the exact nature of the
high-luminosity decline uncertain.

Several surveys have been carried out from the ground over wider areas
to find the rarer, more luminous $z\approx 6$ galaxies. McLure et
al. (2006, 2009) and Curtis-Lake et al. (2012) report on $5<z<6.5$
galaxies found within the 0.8 square degree Subaru/{\it XMM-Newton}
Deep Survey (optical) and UKIDSS Ultra Deep Survey (near-IR). This
SXDS/UDS work provides tighter constraints on the luminosity function
at $-22<M_{1350}<-21$ than from GOODS (McLure et al. 2009). Shimasaku
et al. (2005), Nagao et al. (2007) and Jiang et al. (2011) have
discovered many $z\approx 6$ galaxies in the 0.25 square degree Subaru
Deep Field (SDF) which provides good luminosity function constraints
for $-21<M_{1350}<-20$. Together these works show that the galaxy
luminosity function at $z=6$ can be fit by a Schechter function, the
same parameterization that successfully fits at lower redshift. The
surveys discussed above only contain enough volume to discover galaxies
with a space density of $\gtsimeq 10^{-6}\, {\rm Mpc}^{-1}$. They show
there is a steepening of the luminosity function, but do not
confirm whether the cut-off is exponential or less steep than this.

At even higher redshifts ($z\sim 7 -8$) there have been mixed results
from ground-based surveys to determine the bright end of the
luminosity function. Castellano et al. (2010) found a decrease in the
space density from $z=6$ to $z=6.8$ of a factor of 3.5.  Capak et
al. (2011) found three UV-bright $z\sim 7$ galaxy candidates in the 2
square degree COSMOS field. Although there is some evidence that these
galaxies are low-$z$ interlopers, if they are truly at $z\approx 7$
then the galaxy luminosity function does not decline precipitously as
a Schechter function, but rather as a power-law. Luminous $z\approx 8$
LBG candidates have recently been reported by Yan et al. (2011), also
suggesting a bright end decline less steep than a Schechter
function. Although both the above studies could be affected by low-$z$
contamination, it is important to determine how well the $z=6$
luminosity function is fit by a Schechter function, because there is
little cosmic time available between $z=6$ and $z=7$ for significant
evolution of the shape of the luminosity function. Capak et al. noted
that their result could be explained if AGN feedback is less effective
at early times due to the time required for supermassive black holes
to build up their mass via Eddington-limited accretion and mergers.

The {\it Deep} component of the Canada-France-Hawaii Telescope Legacy
Survey (CFHTLS\footnotemark)
\footnotetext{http://www.cfht.hawaii.edu/Science/CFHTLS} provides the
deepest optical data covering several square degrees and the largest
area survey capable of finding $z \approx 6$ galaxies based on their
rest-frame UV continuum. At a total of nearly 4 square degrees the
volume probed is approximately 40 times that of the GOODS survey and
five times the previous largest area studied, the SXDS/UDS. We present
the results of a search for $z\approx 6$ galaxies in the
CFHTLS. Section \ref{imaging} describes the optical and near-IR data
used and how the galaxies were selected. Section \ref{spectra} gives
details of spectroscopic followup of a subset of the galaxies. Section
\ref{sizes} considers the physical sizes of these galaxies. In Section
\ref{stacked} we stack together the optical and IR images for the
galaxy sample to demonstrate their high-$z$ nature and determine the
typical galaxy SED. Section \ref{lumfun} presents the resulting
luminosity function and our conclusions are drawn in Section
\ref{conc}.

All optical and near-IR magnitudes in this paper are on the AB
system. Cosmological parameters of $H_0=70~ {\rm km~s^{-1}~Mpc^{-1}}$,
$\Omega_{\mathrm M}=0.30$ and $\Omega_\Lambda=0.7$ are assumed
throughout. These parameters have been adopted to ease comparison with
previous work, even though they are slightly different from the best fit
in Jarosik et al. (2011).

\section{Imaging and sample selection}
\label{imaging}

\subsection{Imaging observations}

The imaging data used to select high-redshift galaxies come primarily
from the 3.6\,m Canada-France-Hawaii Telescope. Optical observations
with MegaCam in the $u^*g'r'i'z'$ filters are from CFHTLS Deep which
covered four $\approx 1$ square degree fields with typical total
integration time of 75\,ks in $u^*$, 85\,ks in $g'$, 145\,ks in $r'$,
230\,ks in $i'$ and 175\,ks in $z'$. The seeing in the final stacks at
$i'$ and $z'$ range from 0.66 to 0.76 arcsec. The data used here are
from the 6th data release, T0006, which contains all the data acquired
over the five years of the project\footnotemark
\footnotetext{http://terapix.iap.fr/rubrique.php?id\_rubrique=259}.

These optical data are complemented by near-IR data from the WIRCam
Deep Survey (WIRDS; Bielby et al. 2012). WIRDS used the WIRCam near-IR
imager at the CFHT to cover 2.4 square degrees of the CFHTLS Deep
reaching typical 50\% completeness depth of AB magnitude 24.5 in the
$JHK_s$ filters. A few high-redshift galaxy candidates not in the
regions covered by WIRDS had $J$ band photometry obtained from the
Gemini-North Telescope using GNIRS and from the ESO New Technology
Telescope using SOFI. More recently, near-IR data for some of these
regions has become available from the ESO VISTA telescope. The D1 and
D2 fields are fully covered by the first public data releases of the
VIDEO survey (Jarvis et al. 2012) and UltraVISTA survey (McCracken
et al. 2012), respectively. These data reach about a magnitude
deeper than WIRDS, so all sources in D1 and D2 have $YJHK$ photometry
in this paper from the VISTA surveys instead of the original WIRDS
data. Note that the VIDEO data used has been corrected for the
non-optimal sky subtraction detailed in the release notes. Small parts
of the D2 and D3 field are covered by the \hst Multi-Cycle Treasury
program CANDELS (Grogin et al. 2011; Koekemoer et al. 2011).

Photometry was carried out in dual-image mode using the Sextractor
source extractor software (Bertin \& Arnouts 1996). The $z'$ band was
used as the detection band because it provides the highest
signal-to-noise (S/N) for $z\approx 6$ galaxies. 2 arcsec diameter
photometric apertures were used. Aperture magnitudes were corrected to
total magnitudes assuming that the objects are spatially
unresolved. This gives a lower limit on the flux for spatially
extended objects.

\subsection{Sample Selection}

There are two methods for selecting high-redshift galaxies from
optical/IR broad-band imaging, Lyman break and photometric
redshifts. The Lyman break technique adopts hard color cuts in one or
more colors possibly including non-detections in certain filters,
whereas photometric redshifts use all available filters to derive a
redshift probability distribution. Photometric redshifts are most
suitable when the targets will be detected in many filters. However
they reduce to a Lyman break-type selection if only a few filters are
deep enough to constrain the relevant objects. In this work we use the
Lyman break technique for homogeneous selection because of the
variable near-IR data quality in the different Deep fields.

We set a magnitude limit of $z'<25.3$ to ensure that the objects are
real and not too faint for good photometry. Over most of the survey area this limit corresponds to a $7\sigma$ detection in our 2 arcsec apertures. It will be shown in Section \ref{evidence} that for the fields with the deepest near-IR data (D1 and D2) all 39 $z'$ selected objects (galaxies and brown dwarfs) have a counterpart at $i'$ and/or $J$, showing the rate of spurious $z'$ band detections in the sample is very low. 
The high z' threshold ($\approx 7 \sigma$)  minimizes the problem of photometric
scatter of objects into our sample with true colors different from the
selection criteria. The luminosity function has been well studied by
others at magnitudes fainter than this limit. The primary selection
criterion is color $i'-z'>2$ which corresponds to the break across the
\lya\ line. This criterion is somewhat stricter than other studies
(e.g. Bouwens et al. 2006; Jiang et al. 2011) but ensures that
contamination from low redshift galaxies and brown dwarfs is kept to a
minimum. Two of our objects lie in slightly less deep than average
regions at $i'$ band and are undetected at $i'$ with measured limits
of $i'-z'>1.97$ and $i'-z'>1.99$. These are included in the sample
because there is a high likelihood they would have $i'-z'>2$ if deeper
$i'$ data were available.

Most Lyman break surveys adopt two colour criteria and therefore
define a box in two-dimensional color-color space. We adopt a similar,
but slightly different, method. Instead of a hard cut-off in $z'-J$
color we consider all the properties of every source in the $i'-z'>2$
selection region and determine whether it is most likely a
high-redshift galaxy or something else. This is important here because
the CFHTLS Deep Fields contain varying amounts of extra
multi-wavelength data which can be used as an additional constraint.

\begin{figure}
\vspace{0.3cm}
\resizebox{0.48\textwidth}{!}{\includegraphics{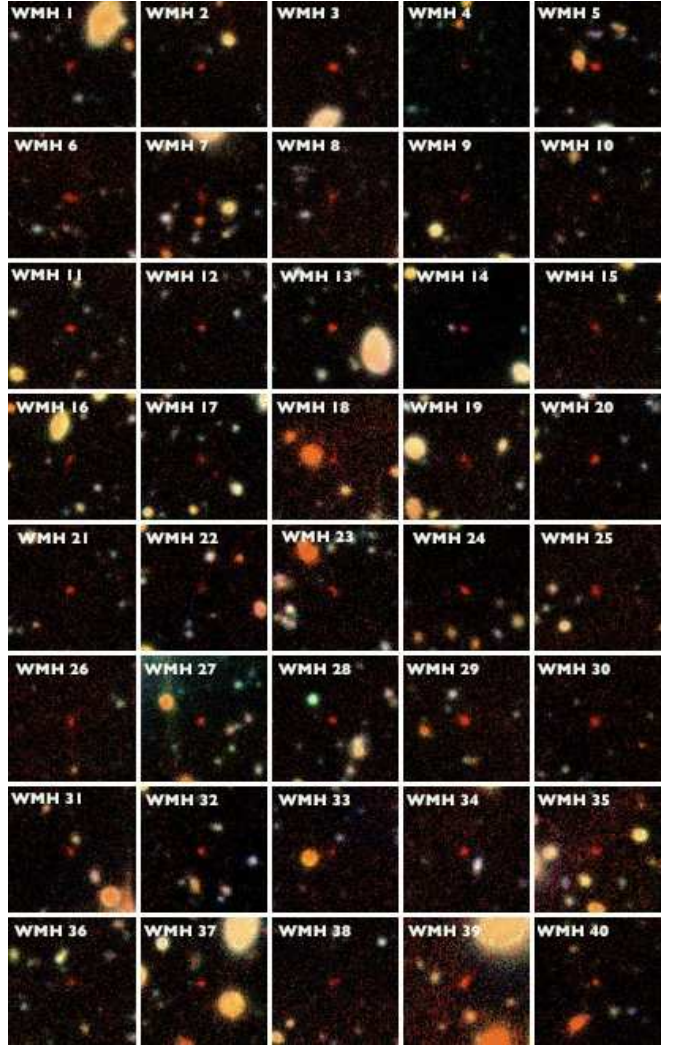}}
\caption{Color images ($r'i'z'$) for the sample of 40 $z\approx 6$
  Lyman break galaxies. Each image covers $15'' \times 15''$ and is
  oriented with north up and east to the left. The LBGs are the small
  red objects at the centers. Some are clearly spatially extended.
\label{fig:colorimages}
}
\end{figure}

The initial automated search routine revealed 136 possible candidates
brighter than the magnitude limit and having $i'-z'>2$ in the $\sim$ 4
square degrees CFHTLS Deep. The images were inspected by
eye. Sometimes more detailed manual photometry for objects in
locations with varying background was performed. Non-detections in the
CFHTLS $u^*g'r'$ filters were also required for good LBG
candidates. We checked that this criterion would not eliminate true
LBGs using the model galaxy simulations to be described in Section
\ref{completeness} that account for the observed variation of galaxy
and intergalactic medium (IGM) properties. It was found that only
0.02\% of galaxies at $5.7<z<6.0$ have colors $r'-z'<3.5$ and hence
could potentially be detected at $r'$ band (typical $2\sigma$ limit is
$r'\sim 27.5$ to $28$) for the brightest galaxies that have $z'
\approx 24.5$.  The most common problems leading to rejection from the
sample were due to structured background near bright stars or close to
the edges of the fields where the effective exposure time is lower. A
total of 69 candidates were removed from the list in this
process. This left 67 true astronomical sources with $i'-z'>2$.

For each of these, the available data was studied to determine its
nature. The primary filter for this process is $J$ because brown
dwarfs are known to have much higher $z'-J$ colors than high-redshift
galaxies. 7 of the 67 sources do not have $J$ band coverage. The next
most important is $Ks$ band because reddened galaxies at $z\approx 1$
would be expected to be bright at $Ks$. In addition, three of the
fields have at least partial \spitzer IRAC coverage available from
SWIRE (Lonsdale et al. 2003), S-COSMOS (Sanders et al. 2007) and AEGIS
(Barmby et al. 2008). The COSMOS field also has deep Subaru imaging
with broad-band depth similar to CFHTLS, additional medium-band
filters (Taniguchi et al. 2007) and HST F814W imaging (Scoville et
al. 2007) which is useful for compact sources.

\begin{deluxetable*}{l l c c c r r c}
\tablewidth{470pt} \tablecolumns{8}
\tabletypesize{\scriptsize}
\tablecaption{\label{tab:photom} Position and photometry of CFHTLS $z=6$ Lyman break galaxies} 
\tablehead{ Name & R.A. and Decl. (J2000.0) &  $i'$ mag  &  $z'$ mag  & $J$ mag      &  $i'-z'$  &  $z'-J$ & FWHM (arcsec)}
\startdata
WMH 1 & 02:24:13.79  $-$04:56:41.4 & $27.18\pm0.26$ & $25.15\pm0.10$ & $> 25.00     $ & $2.02  $ & $< 0.15$ & 1.36 \\
WMH 2 & 02:24:15.10  $-$04:20:47.0 & $27.29\pm0.29$ & $25.14\pm0.10$ & $24.58\pm0.24$ & $2.14  $ & $0.56  $ & 0.92 \\
WMH 3 & 02:24:51.14  $-$04:03:29.4 & $27.42\pm0.32$ & $24.78\pm0.07$ & $24.57\pm0.24$ & $2.63  $ & $0.21  $ & 0.94 \\
WMH 4 & 02:25:19.64  $-$04:28:06.8 & $27.47\pm0.42$ & $25.21\pm0.13$ & $> 25.30     $ & $2.26  $ & $< -0.09$ & 0.88 \\
WMH 5 & 02:26:27.03  $-$04:52:38.3 & $27.47\pm0.34$ & $24.54\pm0.06$ & $24.17\pm0.17$ & $2.92  $ & $0.37  $ & 1.11 \\
WMH 6 & 02:27:18.77  $-$04:50:08.4 & $26.97\pm0.22$ & $24.85\pm0.08$ & $25.16\pm0.39$ & $2.11  $ & $-0.31  $ & 1.94 \\
WMH 7 & 02:27:29.03  $-$04:33:04.3 & $27.69\pm0.48$ & $25.26\pm0.11$ & $24.87\pm0.31$ & $2.42  $ & $0.39  $ & 1.45 \\
WMH 8 & 02:27:46.20  $-$04:30:32.2 & $27.46\pm0.34$ & $25.16\pm0.10$ & $24.35\pm0.20$ & $2.29  $ & $0.81  $ & 1.73 \\
WMH 9 & 09:58:45.49    +02:23:24.8 & $27.67\pm0.49$ & $25.27\pm0.12$ & $24.84\pm0.30$ & $2.39  $ & $0.43  $ & 1.21 \\
WMH 10 & 09:58:59.84   +01:59:48.8 & $27.42\pm0.37$ & $25.24\pm0.12$ & $25.16\pm0.39$ & $2.18  $ & $0.08  $ & 0.84 \\
WMH 11 & 09:59:44.49   +02:09:36.7 & $27.29\pm0.33$ & $25.06\pm0.10$ & $24.90\pm0.32$ & $2.23  $ & $0.16  $ & 1.02 \\
WMH 12 & 09:59:52.74   +02:25:53.2 & $27.34\pm0.41$ & $25.17\pm0.13$ & $25.03\pm0.35$ & $2.17  $ & $0.13  $ & 0.99 \\
WMH 13 & 09:59:56.54   +02:12:27.1 & $26.78\pm0.21$ & $24.78\pm0.08$ & $24.53\pm0.23$ & $2.00  $ & $0.25  $ & 0.86 \\
WMH 14 & 10:00:19.93   +02:25:36.8 & $26.98\pm0.26$ & $24.88\pm0.09$ & $24.83\pm0.07$ & $2.09  $ & $0.05  $ & 0.74 \\
WMH 15 & 10:00:26.37   +02:13:46.8 & $> 27.35     $ & $24.99\pm0.10$ & $24.68\pm0.07$ & $> 2.35$ & $0.31  $ & 1.24 \\
WMH 16 & 10:00:30.58   +02:19:35.1 & $27.56\pm0.42$ & $25.27\pm0.12$ & $24.51\pm0.07$ & $2.28  $ & $0.76  $ & 1.11 \\
WMH 17 & 10:00:50.10   +02:13:03.3 & $27.37\pm0.38$ & $25.17\pm0.12$ & $24.60\pm0.25$ & $2.20  $ & $0.57  $ & 1.05 \\
WMH 18 & 10:00:59.82   +01:57:26.6 & $> 27.27     $ & $25.24\pm0.15$ & $24.63\pm0.26$ & $> 2.02$ & $0.61  $ & 1.10 \\
WMH 19 & 10:01:18.99   +02:11:11.6 & $27.64\pm0.46$ & $25.15\pm0.11$ & $> 25.20     $ & $2.48  $ & $< -0.05$ & 1.29 \\
WMH 20 & 10:01:21.54   +02:12:22.5 & $26.83\pm0.22$ & $24.65\pm0.07$ & $24.14\pm0.17$ & $2.18  $ & $0.51  $ & 0.86 \\
WMH 21 & 10:01:21.94   +02:19:37.6 & $27.20\pm0.30$ & $25.12\pm0.10$ & $24.15\pm0.17$ & $2.07  $ & $0.97  $ & 1.25 \\
WMH 22 & 10:01:38.37   +01:43:48.1 & $27.43\pm0.45$ & $25.28\pm0.13$ & $> 25.20     $ & $2.14  $ & $< 0.08$ & 0.81 \\
WMH 23 & 10:01:38.71   +02:28:20.0 & $> 27.16     $ & $25.05\pm0.12$ & $24.31\pm0.19$ & $> 2.10$ & $0.74  $ & 1.32 \\
WMH 24 & 14:16:20.38   +52:13:23.4 & $> 27.07     $ & $24.72\pm0.07$ & ---            & $> 2.34$ & ---      & 1.03 \\
WMH 25 & 14:17:47.97   +52:38:09.4 & $> 27.28     $ & $25.04\pm0.10$ & $> 25.19     $ & $> 2.23$ & $< -0.15$ & 1.53 \\
WMH 26 & 14:18:10.47   +52:19:43.4 & $27.79\pm0.45$ & $25.13\pm0.10$ & ---            & $2.65  $ & ---      & 1.24 \\
WMH 27 & 14:18:40.04   +52:35:08.2 & $27.20\pm0.32$ & $25.09\pm0.10$ & $> 24.61     $ & $2.10  $ & $< 0.48$ & 1.15 \\
WMH 28 & 14:18:52.23   +53:07:47.3 & $27.31\pm0.29$ & $25.23\pm0.11$ & $> 24.16     $ & $2.08  $ & $< 1.07$ & 1.00 \\
WMH 29 & 14:19:20.53   +52:52:38.7 & $26.52\pm0.14$ & $24.41\pm0.05$ & $24.42\pm0.05$ & $2.10  $ & $-0.01  $ & 1.11 \\
WMH 30 & 14:20:40.80   +52:52:45.3 & $27.47\pm0.34$ & $24.72\pm0.07$ & $> 24.75     $ & $2.75  $ & $< -0.03$ & 1.08 \\
WMH 31 & 14:21:03.74   +52:12:16.2 & $26.90\pm0.21$ & $24.87\pm0.08$ & ---            & $2.02  $ & ---      & 1.08 \\
WMH 32 & 22:13:42.68 $-$17:56:33.2 & $27.42\pm0.38$ & $25.01\pm0.10$ & $> 24.44     $ & $2.41  $ & $< 0.57$ & 0.77 \\
WMH 33 & 22:14:06.97 $-$17:37:59.7 & $> 27.21     $ & $25.24\pm0.13$ & ---            & $> 1.96$ & ---      & 0.73 \\
WMH 34 & 22:14:46.63 $-$17:39:22.2 & $27.31\pm0.35$ & $24.98\pm0.11$ & ---            & $2.33  $ & ---      & 1.02 \\
WMH 35 & 22:15:04.28 $-$17:57:20.5 & $> 27.18     $ & $25.03\pm0.12$ & $> 24.48     $ & $> 2.14$ & $< 0.55$ & 1.44 \\
WMH 36 & 22:15:48.91 $-$17:35:45.7 & $27.30\pm0.34$ & $25.27\pm0.14$ & $> 24.85     $ & $2.02  $ & $< 0.42$ & 1.17 \\
WMH 37 & 22:16:26.67 $-$18:04:45.0 & $> 27.25     $ & $25.26\pm0.13$ & $> 24.12     $ & $> 1.98$ & $< 1.14$ & 0.69 \\
WMH 38 & 22:16:38.04 $-$17:37:00.9 & $27.50\pm0.44$ & $25.21\pm0.14$ & $> 24.85     $ & $2.29  $ & $< 0.36$ & 0.73 \\
WMH 39 & 22:17:08.83 $-$17:38:43.9 & $26.86\pm0.24$ & $24.61\pm0.08$ & $24.57\pm0.26$ & $2.25  $ & $0.04  $ & 1.53 \\
WMH 40 & 22:17:13.55 $-$18:11:45.7 & $27.15\pm0.34$ & $25.03\pm0.10$ & ---            & $2.11  $ & ---      & 0.80 
\enddata
\tablecomments{All magnitudes are on the AB system. $i'$ and $z'$ photometry is from CFHTLS. $J$ photometry is from VIDEO for D1, UltraVISTA for D2 and WIRDS for D3 and D4. Exceptions are  WMH\,14, WMH\,15, WMH\,16 and WMH\,29 which uses CANDELS $J$ data. Spatial size (FWHM) is measured on the CFHTLS $z'$ images.}
\end{deluxetable*}

As discussed in the following section, the results of this process showed that 40 of the 67 candidates were
most likely $z \approx 6$ Lyman break galaxies and the remaining 27 are
most likely to be brown dwarfs.

\subsection{Evidence for $z\approx 6$ galaxies}
\label{evidence}

Three-colour ($r'i'z'$) images for the the 40 candidates classified as
high-redshift galaxies are shown in Figure \ref{fig:colorimages}.
Photometric measurements in the $i'z'J$ filters are given in Table
\ref{tab:photom}. In the case that an object is not detected, a
magnitude limit is given. These magnitude limits are determined from
the variance of photometric aperture fluxes at random locations close
to the object. At $i'$ band, many sources appear to be marginally
detected. Therefore $i'$ magnitudes are quoted down to magnitude errors
of 0.5 magnitudes, rather than the strict $2\sigma$ limits which are
often brighter. This should be kept in mind if using $i'$ magnitudes
or $i'-z'$ colors from Table \ref{tab:photom}. Note that six of the
40 galaxies do not have $J$ band observations. The evidence in favor
of these being high-redshift galaxies will be discussed later. The
final column gives the measured FWHM in the $z'$ band. Many of the
galaxies appear more extended than the seeing, a fact that will be
discussed later.

\begin{figure}
\resizebox{0.51\textwidth}{!}{\includegraphics{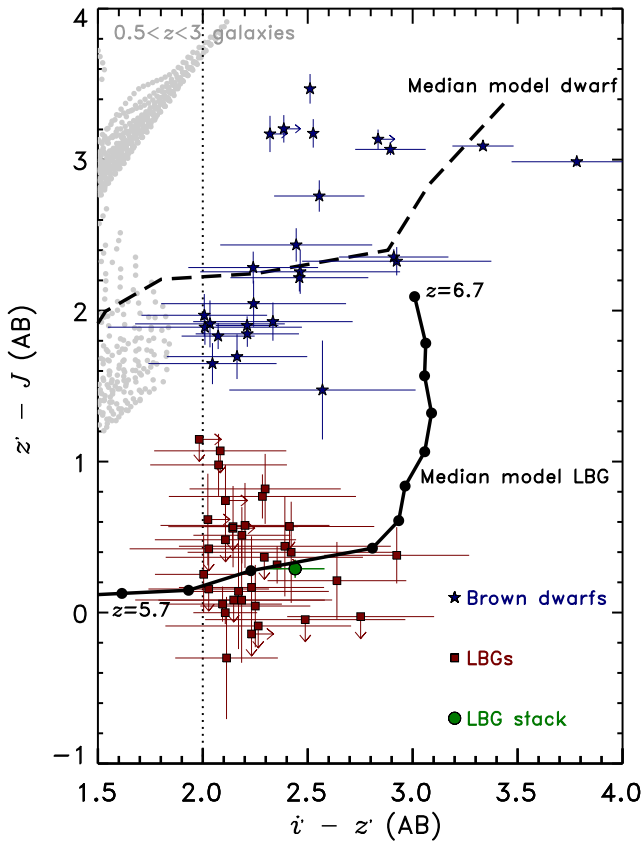}}
\caption{Color-color diagram for the $i'-z'>2$ sources with $i'z'J$ imaging. Those classified as $z\approx 6$ LBGs are plotted as red squares whereas those classified as brown dwarfs are plotted as blue star symbols. The colors of the stacked LBG (Section \ref{stacked}) are shown with a large green circle. The thick black dashed line labeled ``Median model dwarf'' is the expected locus for brown dwarfs based on measurements of closer brown dwarfs. The ``Median model LBG'' curve shows the typical colors expected for LBGs with a range of UV continuum slope and \lya\ emission, as described in Section \ref{completeness}. The gray circles show the possible colors for simulated galaxies at redshifts $0.5<z<3$ (see text for details).
\label{fig:izjcol}
}
\end{figure}

Figure \ref{fig:izjcol} shows the distribution of points on the
$i'-z'$ vs $z'-J$ color-color diagram for the 60 of 67 candidates
which are covered by $J$ band observations. Almost all the sources
detected at $J$ band have colors close to the expected loci of $z>5.7$
Lyman break galaxies or brown dwarfs. The derivation of the brown
dwarf locus is given in Delorme et al. (2008). Close to the $i'-z'=2$
boundary the dwarfs tend to lie below the line. A similar tendency was
noted for the Canada-France Brown Dwarf Survey (Reyl\'e et al. 2010)
and is at least partially due to photometric scatter of somewhat bluer
dwarfs with intrinsic $i'-z'\approx 1.5$. Note that a similar effect
will occur for the high-redshift galaxies. This is why we began with a
relatively high $i'-z'$ cut, so that even those scattered from lower
$i'-z'$ have intrinsic colors matching only brown dwarfs or high-$z$
galaxies. The brown dwarfs in the CFHTLS/WIRDS survey also have
well-defined loci in the $z'-J$ vs $J-H$ and $J-H$ vs $H-Ks$ diagrams
with no significant outliers.

The gray circles show colors of model galaxies at $0.5<z<3$ determined
using the 2008 stellar population synthesis models of Charlot \&
Bruzual (priv. comm.). Passively evolving and dusty star-forming
galaxies with ages ranging from a million years to the age of the
universe at that redshift are included. Extreme dust extinction with
optical depth up to $\tau=10$ is included, because extreme dust is
required to get such large $i'-z'$ colors. Note that due to the larger
wavelength difference between $z'$ and $J$ than between $i'$ and $z'$,
the effect of extreme dust reddening is to push low-$z$ galaxies into
the brown dwarf region of the diagram. A combination of extreme
reddening plus high photometric scatter would be required to obtain
the colors of our sample. The low-$z$ galaxy models closest to the
location of the LBGs have $z \approx 1.2$ and $\tau \approx 10$. Such
galaxies would be extremely rare and faint.

21 of the 34 LBG-classified targets with $J$ band data are detected at
$J$. The other 13 have $z'-J$ limits well separated from the regions
covered by brown dwarfs and reddened $0.5<z<3$ galaxies. We expect
almost all the red squares in Figure \ref{fig:izjcol} to be
high-redshift galaxies as discussed below.

As noted in Table 1, many of the LBGs appear spatially resolved at
$z'$ band. Section \ref{sizes} discusses this in much more detail and
it is concluded that at least half of the LBGs are spatially resolved
from the ground. In addition, \hst observations of a subset also show
that they are resolved. This provides further evidence against
contamination by brown dwarfs.

A final source of contamination to consider is that by quasars at
$z\approx 6$ which will have similar colors. A search for such quasars
going to magnitude $z'<24.5$ found none in the CFHTLS Deep survey area
(Willott et al. 2010), but one at $z=6.01$ with magnitude $z'=24.4$ in
the SXDS/UDS (McLure et al. 2006; Willott et al. 2009).  At such faint
magnitudes, the quasar luminosity function slope is likely to be
fairly flat (Willott et al. 2010), so the expectation would be for
$\approx 1$ more quasar in the magnitude range $24.5<z'<25.3$ across
the full survey region.

\subsection{Sources without $J$ photometry}

7 of the 67 $i'-z'>2$ sample do not have $J$ photometry because they
lie outside the WIRDS and VISTA surveys and their nature, high-$z$
galaxy or brown dwarf, had to be ascertained using other available
information. Given the ratio of galaxies to brown dwarfs in the regions 
with $J$ photometry we would expect 4 of these to be galaxies and 3 
to be brown dwarfs.

One object is in the D4 field and a WIRCam $Ks$ band image
provided by Genevieve Soucail shows it to have $Ks=22$. This red color
of $z'-Ks=2.8$ is consistent with a L dwarf.

The six remaining objects are retained in the list of likely $z
\approx 6$ galaxies. WMH\,24, WMH\,26, WMH\,31 and WMH\,34 have measured
$z'$ band FWHM $> 1$ arcsec and therefore appear to be spatially
resolved. As will be discussed in Section \ref{sizes}, only 15\% of
brown dwarfs are found to have such high FWHM. The probability that
none of these four are brown dwarfs based on the FWHM argument is
therefore 60\%.  WMH\,33 has FWHM=0.73, so is consistent with being
unresolved. However it is faint at $z'$ band with $z'=25.25$ and most
sources close to the detection limit are galaxies. It is not detected
in a WIRCam $Ks$ band image that reaches to $Ks=22$. If it were a L
dwarf it would be expected to have $Ks \approx 22$ to $22.5$, so this
limit, although not definitive, suggests that it may be a galaxy.
Therefore it is retained in the galaxy sample. WMH\,40 has FWHM=0.80
arcsec and $z'=25.05$. It is retained within the galaxy sample but we
note it has almost equal probability to be a high-$z$ galaxy, or a
brown dwarf with these properties. In conclusion, at least four of
these six objects are likely high-$z$ galaxies. Taking into
consideration there could be one $z\approx 6$ quasar in our sample,
the contamination of our $z\approx 6$ galaxy sample is most likely
$<10$\%.

\section{Spectroscopy}
\label{spectra}

\subsection{Observations}

The photometrically-selected high-$z$ galaxies are spread across
nearly 4 square degrees. Given that the typical field-of-view of
multi-object spectrographs on large telescopes is $\ll 1$ degree,
there is little opportunity for a high-multiplex factor. Spectroscopic
observations have been attempted for 7 of the 40 candidates. Priority
was given to those which are brightest at $z'$ band to minimise the
integration time required. All spectroscopy was performed using the
GMOS spectrographs on the 8.2\,m Gemini Telescopes. The R400 gratings
were used with 1 arcsec slits to give a resolution of $R=1000$. CCD
pixels were binned by a factor of two in the spectral direction so that
each binned pixel covers 1.34\,\AA. WMH\,13 was observed in long-slit mode, WMH
29 was observed equally in long-slit and multi-object mode, and all
other observations used multi-object mode. Occasionally, two $z\approx
6$ galaxies could be fit on the same mask. The masks were filled with
lower redshift Lyman break galaxies and photometric redshift selected
galaxies ($z>3.5$). All observations used the nod-and-shuffle mode to
ensure good subtraction of the sky background spectrum.

\begin{deluxetable}{l c c c c c c}
\hspace{-0.4cm}
\tablewidth{240pt}
\tablecolumns{7}
\tablecaption{\label{tab:spec} GMOS spectroscopy of $z\approx 6$ Lyman break galaxies} 
\tablehead{Name & Integ.   & $M_{1350}$ &  $z_{\rm spec}$ &  QF$^{\rm a}$ & $W_{{\rm Ly}\alpha}$$^{\rm b}$ & ${\rm HWHM}_{{\rm Ly}\alpha}$$^{\rm c}$ \\
                & (hr)  &         &                &     & (\AA)        & (km\,s$^{-1}$)}
\startdata
WMH 5    & 2.5  &  $-22.65$ & $6.068$  & A & $13 \pm 4$ & $220 \pm 30$ \\ 
WMH 6    & 2.5  &  $-21.75$ & $5.645$  & C & --         & -- \\ 
WMH 13   & 2.5  &  $-22.06$ & $5.983$  & A & $27 \pm 8$ & $210 \pm 50$ \\ 
WMH 15   & 4.0  &  $-21.98$ & $5.847$  & C & --         & -- \\ 
WMH 29   & 5.0  &  $-22.27$ & $5.757$  & B & $4 \pm 2$ & unres.? \\ 
WMH 34   & 2.3  &  $-21.67$ & $5.759$  & C & --         & -- \\ 
WMH 39   & 2.0  &  $-22.06$ & $5.733$  & B & --         & -- \\ 
Serendip & 2.5  &  $-21.33$ & $5.618$  & A & $31 \pm 11$ & $240 \pm 40$
\enddata
\tablenotetext{a}{QF is the Quality Flag of the redshift determination. See text. Note that QF=C redshifts are uncertain and $z_{\rm spec}$ merely gives the best fit redshift.}
\tablenotetext{b}{\,Ly\,$\alpha$ rest-frame equivalent width, quoted only if \lya\ is detected. Uncertainties in $W_{{\rm Ly}\alpha}$ are dominated by continuum placement.}
\tablenotetext{c}{Half-width half-maximum of the \lya\ line measured from the peak to the red side of the line. These values have been deconvolved for the instrumental dispersion. The measured ${\rm HWHM}_{{\rm Ly}\alpha}$ for WMH\,29 is $150 \pm 60$, which is equal to the instrumental dispersion, but with significant uncertainty due to its low equivalent width.}

\end{deluxetable}

\begin{figure}
\resizebox{0.49\textwidth}{!}{\includegraphics{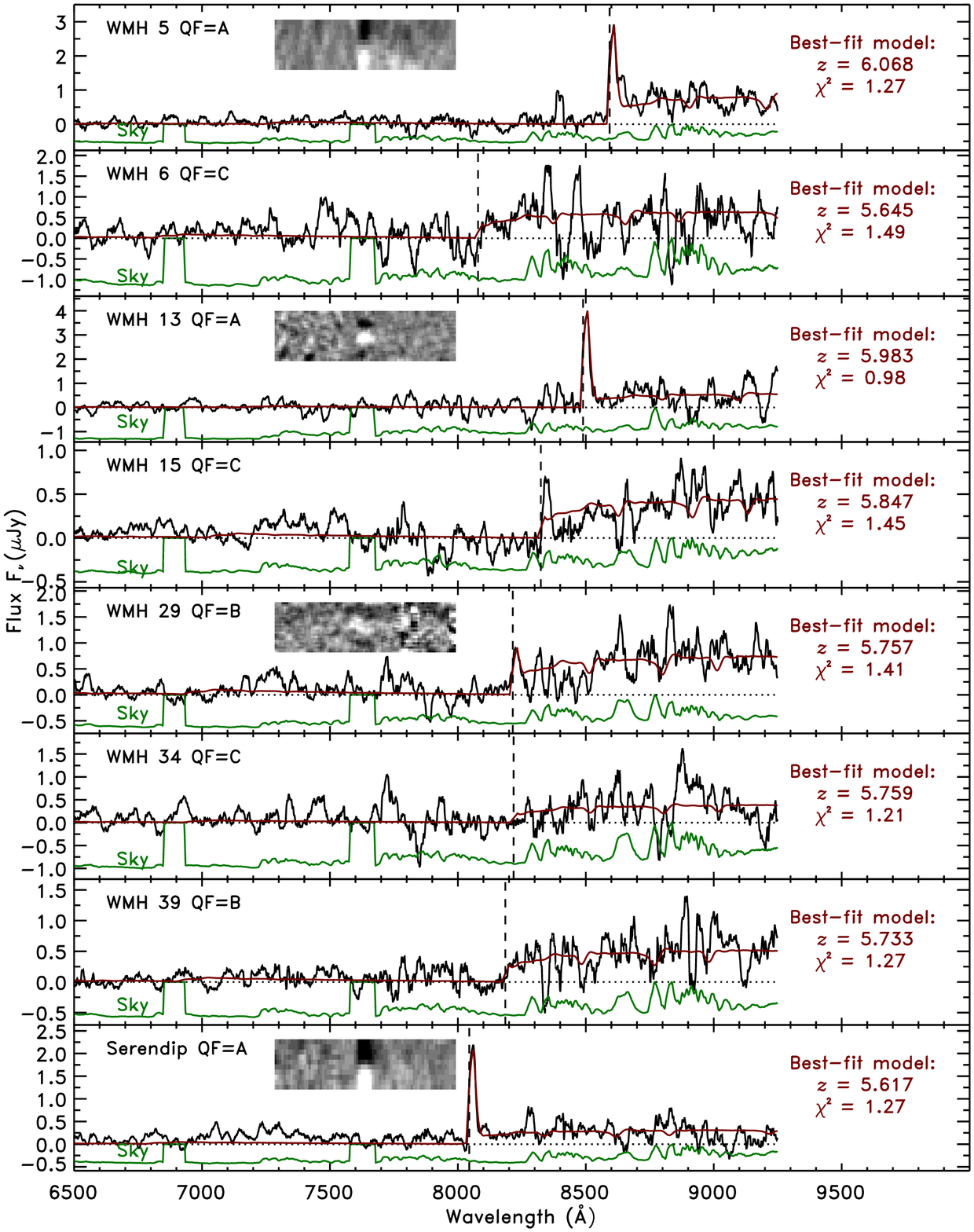}}
\caption{Optical spectra of $z\approx 6$ Lyman break candidates
 (black) with the best fitting galaxy models (red). The best-fit
 redshifts and reduced $\chi^2$ are quoted. All spectra are smoothed
 by 20 pixels (27 \AA). Note that template fitting was performed on
 the unsmoothed spectra. The noise (dominated by the sky) is plotted
  in green and offset from zero. The noise is not to scale, but
  included to show when observed spectral features may be due to
  noise. The dashed lines mark the wavelengths of the \lya\ transition
  for the best-fit redshifts. For the four LBGs with \lya\ emission the inset panel shows a 220 \AA\ long portion of the 2D spectrum centred on the \lya\ emission. Due to nod-and-shuffle observing mode, these spectra show positive and negative versions offset vertically. WMH\,5 and Serendip were observed in multi-slit mode so there is one positive (white) below one negative (black). WMH\,13 and WMH\,29 were observed in long-slit mode so there is one positive (white) with two negative (black) above and below.
 \label{fig:spec}
}
\end{figure} 

The long-slit observations were performed in queue-mode in excellent
conditions in 2009. The multi-object observations were performed in
classical mode on the nights of 12 December 2010, 27 and 28 June 2011
in variable conditions. Details of the spectroscopic observations and
results are given in Table \ref{tab:spec}. Due to the faintness of
these targets, the typical continuum S/N per pixel between
8500\,\AA\ and 9000\,\AA\ was only $\sim 0.5$. Therefore, redshifts could
only be quickly and unambiguously determined for the few sources which
showed very strong \lya\ emission lines. However, even at S/N $=0.5$
per pixel, the continuum can be significantly detected over broader
wavelength intervals and redshifts determined from continuum breaks
(Spinrad et al. 1998).

\subsection{Model fitting}
 
To fully utilize the spectral information, a routine was developed to
fit the observed spectrum to model templates. The galaxy models used
are based on the observed composite $z=3$ Lyman break galaxy spectrum
of Shapley et al. (2003). Two galaxy models are used: one with a
strong \lya\ emission line and one with only \lya\ absorption. The
fitting routine allows a combination of these models to represent a
range in \lya\ strength. IGM absorption is accounted for using the
model of Songaila (2004). The galaxy spectrum is multiplied by a
power-law to account for the variation in observed UV continuum slope
of galaxies. There are a total of four free parameters: normalization,
redshift, power-law slope and \lya\ flux. The \lya\ flux was fixed at
zero for those galaxies without obvious \lya\ emission in their 2D
spectra to avoid over-fitting of noise peaks with the \lya\ line.

Best-fit galaxy models are determined by fitting the models to the
observed spectra. The uncertainty on each spectral pixel due to the
sky noise is determined from the variance of blank sky pixels in each
of the masks. The best fit is determined by the lowest value of the
reduced $\chi^2$. The spectra, along with best fit galaxy models, are
shown in Figure \ref{fig:spec}. The observed spectra, models and sky
noise spectrum have been smoothed by 20 pixels (27\,\AA) for display
purposes. Those sources with measurable \lya\ emission have an inset
panel in Figure \ref{fig:spec} with a 220\,\AA\ long segment of the 2D
spectrum centred on the \lya\ emission.

\subsection{Results of spectroscopy}
 
Each target has a redshift {\it Quality Flag} in Table \ref{tab:spec}
which gives a measure of the confidence of the redshift
determined. The categories follow those in Vanzella et al. (2009)
where `A' means unambiguous, `B' is likely and `C' is uncertain.
Strong \lya\ emission was only found in two of the $z\approx 6$
galaxies, WMH\,5 and WMH\,13. These are assigned QF=A. In addition,
weak, but still significant, \lya\ was observed in WMH\,29 which is
assigned QF=B. Note that this \lya\ line at $z=5.757$ is in a region
of very low sky noise. It is doubtful that such a line would have been
significantly detected at slightly higher redshift. Due to the
weakness of this line it cannot be determined if it is asymmetric and
therefore definitely \lya. If it is instead \oii\ at $z=1.20$ then the
4000\,\AA\ break would be expected at 8790\,\AA, whereas the continuum break is
observed to be at $<8500$\,\AA.  The apparent emission line at
8400\,\AA\ in WMH\,15 in Figure \ref{fig:spec} is found to be due to
sky lines upon examination of the full resolution spectrum.

One of the serendipitous mask-filling galaxies with photometric
redshift $z>3.5$ was also found to have strong \lya\ at $z=5.618 $,
just below the redshift range of our sample. It is labeled
``Serendip'' and is the last object in Table \ref{tab:spec} and Figure
\ref{fig:spec}. It has QF=A and is located at 02:26:37.02
$-04$:55:20.4. This galaxy has $i'-z'=1.1$, well below our selection
criterion of $i'-z'=2.0$. The reason for this much lower color is that
the strong \lya\ line is still within the $i'$ filter at this
redshift. When an emission line of this strength shifts from the $i'$
band to the $z'$ band (at $z\approx 5.8$) it increases the $i'-z'$
color by 0.5.

All four of the galaxies with \lya\ emission show significant spectral
breaks across the \lya\ line. All four have best fit galaxy spectra at
redshifts equal to their \lya\ redshifts. This is not too surprising
for those three with strong \lya\ lines, because inability to fit the
line would increase the $\chi^2$. However it is encouraging for WMH
29, which has only weak \lya, where the continuum break contributes
most of the weight in the fit. For the four galaxies without \lya\
emission, we depend upon continuum breaks to constrain the redshift.
WMH\,39 also shows a break at $z=5.73$ and is assigned QF=`B' based on
this break. WMH\,15 shows a likely continuum break between $z=5.85$
and $z=5.95$, however this break is not so clear as that in WMH\,39 leaving the redshift uncertain and hence QF=`C'. 
 
The other two galaxies, WMH\,6 and WMH\,34, show continuum flux in the
$z'$ band and regions with no flux at shorter wavelength, but the
exact break redshifts are not well-constrained, so are assigned
QF=`C'. These two spectra have the highest noise (along with WMH\,13
which has the strongest \lya\ line) which likely accounts for the lack
of a clear redshift solution. The best fit redshift of $z=5.645$ for
WMH\,6 is quite unlikely given its color of $i'-z'=2.1$ (see Section
\ref{completeness}). We note that in any color-selected sample there
is a possibility for rare interlopers with unusual colors (e.g. Capak
et al. 2011; Hayes et al. 2012) to masquerade as high-redshift
galaxies. However, based on the evidence we have and the stringent
color cuts, we expect there to be very few of these in our
sample. Only WMH\,13 out of the seven color-selected spectroscopic
targets has a \lya\ line with rest-frame equivalent width $>25$\,\AA\
giving a fraction of 14\%. This is comparable to the fraction of $20
\pm 8$\% for luminous $z\sim 6$ LBGs found by Stark et al. (2011) and
lower than the $54 \pm 21$\% observed by Curtis-Lake et al. (2012). 

Table \ref{tab:spec} includes the absolute magnitudes, $M_{1350}$ of
the galaxies. These were calculated from the $z'$ band magnitudes
after taking account of the observed \lya\ emission line
contributions. At these redshifts, the $z'$ band contains flux at or
very close to rest-frame 1350\,\AA\ and hence the main uncertainty on
the absolute magnitudes comes not from $k$-corrections, but the $z'$
band photometry and is $\approx 10$\%. With absolute magnitudes
ranging from $-21.67$ to $-22.65$, the spectroscopic targets are some
of the most luminous LBGs known at $z\sim 6$. For comparison, the
break in the $z\approx 6$ luminosity function is at $M_{1350}=-20.2$
(Bouwens et al. 2007). WMH\,5 has $M_{1350}=-22.65$, which is even
more luminous than the lowest known luminosity $z\approx 6$ quasar,
CFHQS\,J0216-0455 ($M_{1450}=-22.21$, Willott et al. 2009). WMH\,5 is
only 2.1 arcsec from the center of a galaxy in Figure
\ref{fig:colorimages}. This is an inclined disk galaxy with a
photometric redshift of $z=1.01$, rest-frame $B$-band absolute
magnitude of $M_B=-21.5$ and disk half-light radius $4$\,kpc.  Using
the results of studies of the Tully-Fisher and similar relations for
$z=1$ disk galaxies (Dutton et al. 2011; Miller et al. 2011), we find
the circular velocity for this galaxy should be $\approx
160$\,km\,s$^{-1}$. Modeling the gravitational potential as an
isothermal sphere, the gravitational lensing magnification of WMH\,5
due to this galaxy is a factor of 1.27. Hence the intrinsic absolute
magnitude of WMH\,5 is still extremely luminous with $M_{1350} \approx
-22.4$.

\section{Galaxy sizes}
\label{sizes}

An important cosmological observation is that galaxies are smaller at
higher redshifts than similar galaxies at lower redshift (Ferguson et
al. 2004; Bouwens et al. 2004). This is expected due to the evolution
in the virial radii of dark matter halos, but the exact nature of the
evolution also depends upon details of gas accretion, retention and
star formation efficiency. The faint $z\sim6$ galaxies discovered in
GOODS and HUDF typically have half-light radii of $\approx 0.1$ arcsec
(equivalent to 0.6\,kpc; Bouwens et al. 2006) and are therefore not expected to be
resolved in typical seeing-limited, ground-based observations. Table
\ref{tab:photom} showed that the measured $z'$ band FWHM of many
CFHTLS $z\approx 6$ LBGs are larger than the seeing of $\approx 0.7$
arcsec of the images.

\begin{figure}
\resizebox{0.48\textwidth}{!}{\includegraphics{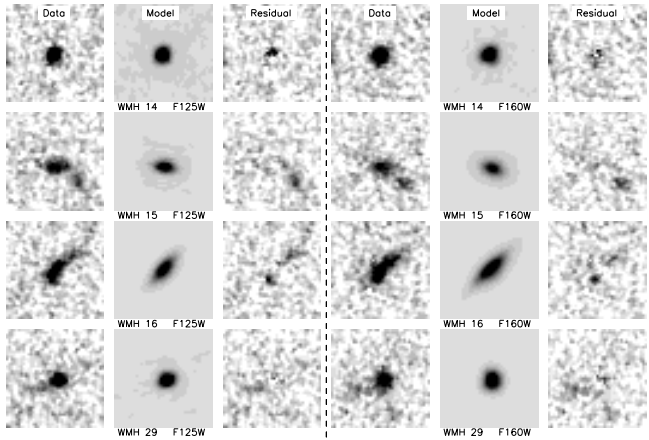}}
\caption{WFC3 F125W (left) and F160W (right) imaging of the four LBGs in the CANDELS survey. For each observation there are three panels shown from left to right: (i) the data; (ii) the {\sc GALFIT} best fit single galaxy model; (iii) the residuals after subtraction of the smooth galaxy profile. All images are 2.4 arcsec across, corresponding to a linear physical size of 14\,kpc. The images of WMH\,29 are oriented at an angle of 50$^\circ$ counter-clockwise to the standard north-up, east-left orientation of the other three galaxies. For all galaxies, there are significant residuals with similar structure in the two filters. In the case of the lower three galaxies these residuals reveal multiple components interacting with the main galaxies.
\label{fig:wmhgalfit}
}
\end{figure}

Four (WMH\,14, WMH\,15, WMH\,16 and WMH\,29) of the 40 LBGs (including
two with spectroscopy) are located within the CANDELS survey and
therefore have high-resolution \hst ACS and WFC3 imaging available
(Grogin et al. 2011; Koekemoer et al. 2011). We have retrieved these
data in order to get a more detailed view of the galaxy morphologies
for this small sub-sample. The WFC3 F125W and F160W images are shown
in Figure \ref{fig:wmhgalfit}.  We have used {\sc GALFIT} (Peng et
al. 2010) to fit galaxy models to the main galaxy component in each
filter independently. The best fit single model galaxy and the
residuals after subtraction are also plotted in Figure
\ref{fig:wmhgalfit}. Similar models and residuals are found for both
filters in all cases, providing a high degree of confidence in the
model galaxy fits. Inspection of the F814W images showed that the
residuals have similarly red F814W-F125W colors as the main galaxies,
indicating the residuals are at $z\approx 6$ and not at lower
redshift.

For WMH\,14 the residual emission is compact and co-incident with the
main galaxy. For the other three galaxies, the residual emission is
more extended and clumpy, extending at least an arcsec (equivalent to
6\,kpc) from the main galaxy centroid. This is suggestive of galaxy
mergers and interactions being common in the most luminous galaxies at
this epoch. The model galaxies have a wide range of half-light radii
from only 0.3 and 0.5\,kpc for WMH\,14 and WMH\,29, respectively, to
1.0 and 1.5\,kpc for WMH\,15 and WMH\,16, respectively.

Six of the 15 galaxies in D2 are well-detected in the ACS F814W
observations of the COSMOS field. They are all spatially resolved and
half of them show multiple components. The F814W filter extends
further redward than the CFHT $i'$ filter, so such detections in F814W
are consistent with the $z\approx 6$ LBG classification. The frequent
occurrence of multiple components in both the WFC3 and ACS data
indicate that about half of the LBG sample have signs of
interactions/mergers.

\begin{figure}
\vspace{-0.4cm}
\resizebox{0.51\textwidth}{!}{\includegraphics{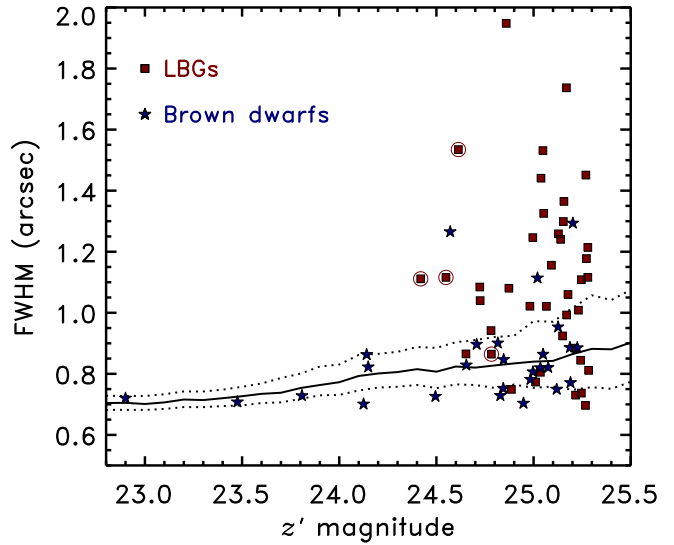}}
\caption{Observed size of the selected Lyman break galaxies (red squares) and brown dwarfs (blue stars).  It is evident that the Lyman break galaxies tend to have larger sizes than the unresolved brown dwarfs. The solid line shows the median measured FWHM for artificial point sources added to the images in our completeness simulations (see Section \ref{completeness}). The typical measured FWHM of point sources increases at faint magnitudes due to noise. The 25\% and 75\% quartiles are shown with dotted lines. The distribution of brown dwarfs is consistent with these lines, whereas about 50\% of the LBGs are truly spatially resolved. The spectroscopically confirmed LBGs (QF=`A' or `B'; Section \ref{spectra}) are enclosed by a larger circle. Three of these four are spatially resolved.
\label{fig:size}
}
\end{figure}

The only size data available for the full sample are the measured FWHM
in the $z'$ band.  Figure \ref{fig:size} plots FWHM against magnitude
for the 40 LBGs and 27 brown dwarfs. The brown dwarfs are included as
a comparison sample, because they are expected to be almost all
unresolved. Although many brown dwarfs exist in binary systems
(Burgasser et al. 2007) only a small fraction would have the relevant
component separations to appear as a single, resolved source in these
images. The spread in observed FWHM for dwarfs is due to three
factors: the small difference in quality between the four Deep fields,
the variation in stellar FWHM from the field centers to the edges and
photometric noise. The photometric noise is evident by the increased
spread in FWHM for dwarfs at fainter magnitudes. It is for this reason
that we use the dwarfs as the unresolved comparison sample to compare
to the LBGs rather than stars at brighter magnitudes.

Figure \ref{fig:size} also shows the 25\%, 50\% and 75\% centiles of
the measured FWHM distribution as a function of magnitude for
simulated point sources inserted into the data (see Section
\ref{completeness}).  The observed distribution for the dwarfs agrees
well with the simulation. Note that the three dwarfs with measured
FWHM $>1$ arcsec all have FWHM $\approx 0.6$ in $J$ band
imaging with higher S/N than $z'$ band for these red sources,
confirming photometric noise as the reason for a small fraction of
large FWHM values. It is found that 3 out of 20 (15\%)  dwarfs with
magnitudes in the range of LBGs are observed to have FWHM $>1$ arcsec
and lie beyond the 75\% simulated centile due to noise. In contrast, we find
that 26 out of the 40 LBGs (63\%) have FWHM $>1$ arcsec and that the
same percentage lie beyond the 75\% centile. Assuming that 15\% of
these LBGs are outliers due to noise, then there remain $\approx 50$\%
which are truly resolved in our ground-based imaging. 

Given the uncertainty in the broadening of the FWHM values by seeing
and noise, it is not possible to reliably deconvolve each value to
determine the intrinsic size distribution of our LBG sample. However,
a typical intrinsic half-light radius of 2\,kpc (0.35 arcsec) is
required to observe a galaxy with FWHM 1.1 arcsec (median for our LBG
sample) when the FWHM of a point source would be 0.85 arcsec (50\%
centile for the simulated point sources with magnitude
$z'=25.0$). This contrasts with the typical half-light radius of
1\,kpc for $z=26$ galaxies in GOODS (Bouwens et al. 2006). The
high-resolution WFC3 imaging of four LBGs showed that one is compact,
one has a compact core with low surface brightness extended emission
and two have relatively large half-light radii plus extended
emission. This suggests that both large galaxy sizes and multiple
emission components are common in our sample of the most luminous
LBGs. Our finding is consistent with an extension to the brightest
galaxies of the strong correlation between UV luminosity and linear
size at $z=6$ presented by Dow-Hygelund et al. (2007).

\begin{figure*}
\resizebox{1.00\textwidth}{!}{\includegraphics{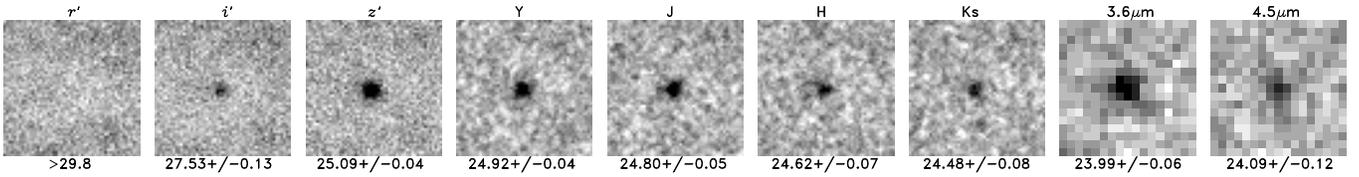}}
\caption{Median stacked images in the $r'i'z'YJHKs$ and IRAC 3.6$\mu$m and 4.5$\mu$m filters at the positions of the LBGs used to construct the stacks. There are 23 galaxies in the optical (CFHTLS) and $YJHK$ (VIDEO and UltraVISTA) stacks and 15 in the IRAC bands (S-COSMOS). Each image covers $10'' \times 10''$. The images are oriented with north up and east to the left. Negative holes in the central few arcsec of the images (particularly apparent in $r'$ and $i'$) are due to incompleteness of our LBG sample due to foreground contamination.
\label{fig:imstack}
}
\end{figure*}

\begin{figure}
\hspace{-0.4cm}
\resizebox{0.51\textwidth}{!}{\includegraphics{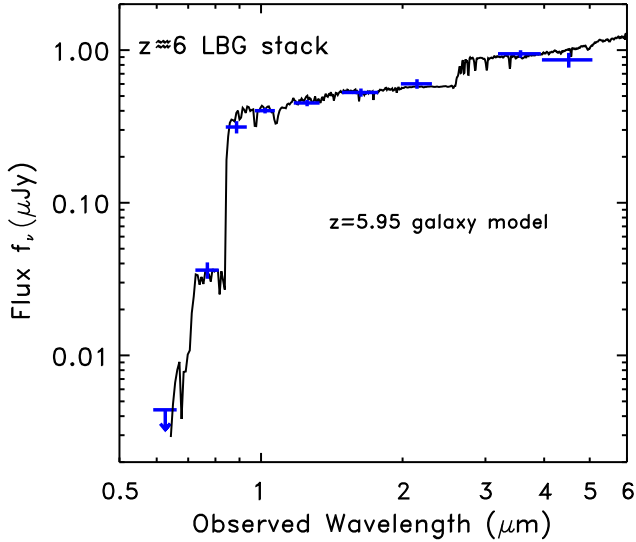}}
\caption{Photometry from the stacked images (blue points with error bars) plotted with stellar
  evolution synthesis model (black line). The $z'$ band flux has been reduced by a factor of 1.1 to
  account for the typical \lya\ emission, which is not included in the models. The galaxy model has constant star formation rate, solar metallicity and a Chabrier IMF. The best-fit parameters are redshift $z=5.95$, stellar mass $=1.1\times 10^{10}\,{\rm M}_\odot$,  age $=1.3\times  10^{8}$\,yr, star formation rate $= 110 \,{\rm M}_\odot{\rm yr}^{-1}$ and reddening $A_{V}=0.75$. See text for further details.
\label{fig:sedstack}
}
\end{figure}

\section{Stacked galaxy properties}
\label{stacked}

In order to check that the typical LBG properties match those of
high-redshift galaxies and to get an idea of the typical galaxy
spectral shape, we have stacked the photometric data at the positions
of the $z \approx 6$ galaxies. After registering the images, the
median flux of each pixel was determined since it provides the most
robust estimator that is least affected by outliers (White et
al. 2007). Because the near-IR data from VISTA has an extra filter
($Y$) and is deeper than WIRDS, we just use the 23 galaxies in D1 and
D2 with VISTA data. We have checked that the stacked optical and
near-IR magnitudes for D3 and D4 are consistent. Deep \spitzer IRAC
data is available for the 15 galaxies in D2 from the S-COSMOS survey
(Sanders et al. 2007). We include these data because these wavelengths
are important for constraining the stellar mass and age, but the
reader should remember that the IRAC points only correspond to a
subset of the objects used for the $r'i'z'YJHKs$ stacks.

The resulting median stacked images are shown in Figure
\ref{fig:imstack}. There are significant detections at all filters
except for $r'$ band. Photometry of the ground-based data was
performed to obtain 3 arcsec aperture magnitudes or limits. These are
equivalent to total magnitudes, within the uncertainties. These 3
arcsec apertures are larger than the 2 arcsec used previously in this
paper for individual object magnitudes (many of which had low S/N
necessitating small apertures) and provide a better match to the IRAC
apertures. For the IRAC data which has a broader PSF, 3.8 arcsec
apertures were used and a PSF correction to total magnitudes
applied. Smoothing of the images showed that for all the optical
filters, there is a negative ``hole'' in the background covering the
central five arcseconds. This is most apparent for $r'$ and $i'$ in
Figure \ref{fig:imstack}. This is showing us that our $z=6$ galaxy
sample is incomplete. The CFHTLS Deep $r'$ and $i'$ band data is so
deep that it is close to the confusion limit and we are missing $z=6$
galaxies that have foreground galaxies contaminating the photometry
aperture in $r'$ and $i'$. Another consequence of this is that we are
unlikely to have strongly gravitational lensed galaxies in our sample
due to the lensing galaxy contaminating the source galaxy
aperture. Photometric measurements were performed carefully to ensure
that the appropriate background level was set. The measured magnitudes
of the stacks are given below the images in Figure
\ref{fig:imstack}. The 2$\sigma$ limit on the $r'$ band magnitude
(determined from the noise in the background pixels because this
region has lower variance than the rest of the field which was
populated by galaxies in the input images) gives $r'>29.8$. This is
almost 5 magnitudes fainter than the $z'$ band magnitude and provides
compelling evidence that these galaxies are truly at $z\approx 6$
rather than lower redshift galaxies or dwarf stars. The stack colors
of $i'-z'=2.44 \pm 0.14$ and $z'-J=0.29 \pm 0.06$ are plotted in
Figure \ref{fig:izjcol} and are entirely consistent with a typical
$z=5.9$ galaxy.

Figure \ref{fig:sedstack} plots the stack fluxes as a function of
wavelength. The $z'$ flux has been reduced by a factor of 1.1 to
account for the typical \lya\ emission line contribution. It is
apparent that the $YJHKs$ fluxes are described by a power-law with
index (defined as $f_\lambda \propto \lambda^{\beta}$) redder than
$\beta=-2$ (which would be flat on this plot which uses $f_\nu$). A
power-law fit to the $YJHKs$ fluxes gives $\beta=-1.44 \pm 0.10$,
where the $1\sigma$ uncertainty comes from 1000 bootstrap resample
trials. To ensure that this result is not an artifact of the stacking
method, we have also fitted the $YJHKs$ fluxes of the 23 LBGs
individually. Excluding two extreme outliers which have poor fits, it is
found that the mean $\beta=-1.38$ and the median $\beta=-1.02$. The
standard error on the mean is $\pm 0.20$.

The value of $\beta$ found is significantly redder than the typical
values previously observed for the most luminous $z\approx 6$ LBGs in
deep \hst surveys of $\beta=-1.78 \pm 0.11$ (Bouwens et al. 2012),
$\beta=-2.10 \pm 0.16$ (Dunlop et al. 2012), $\beta=-2.04 \pm 0.17$
(Wilkins et al. 2011) and $\beta=-2.05 \pm 0.11$ (Finkelstein
et al. 2012). Finkelstein et al. showed that although at $z\approx 6$
there is only a weak correlation between $\beta$ and UV luminosity,
there is a stronger correlation between $\beta$ and stellar
mass. Whilst this is not surprising, because in a UV flux-limited
sample, a redder galaxy model will require a higher stellar mass, the
typical value of $\beta$ for their most massive bin ($10^9$ to
$10^{10}$M$_\odot$) is $-1.78$ and the typical stellar mass of our
sample is $10^{10}$M$_\odot$ (see below). The simplest explanation for
dust in luminous galaxies comes from combining the metallicity--dust,
mass-metallicity and mass-luminosity correlations. Therefore our
typical $\beta=-1.4$ for $z\approx 6$ luminous LBGs is not at odds
with the work of Finkelstein et al. Our findings suggest that a
substantial number of these galaxies have significant dust reddening.

Galaxy models were fitted to the stacked photometry using the
photometric redshift method of McLure et al. (2011). The fitted models
used Bruzual \& Charlot (2003) stellar populations with the Chabrier
IMF and the Calzetti et al. (2000) dust attenuation law. Various star
formation histories were fitted, but none were preferred at a high
significance. Therefore we quote the constant star formation rate
results, since these most-luminous LBGs are unlikely to be observed in
a state of rapidly declining or increasing star formation rate. Figure
\ref{fig:sedstack} shows the best-fit model and its parameters. The
best fit redshift was found to be $z=5.95$ with a $2\sigma$ range of
$5.87<z<5.97$. The best-fit metallicity is solar. The best-fit stellar
mass, $M_\star$, is $1.1\times 10^{10}{\rm M}_\odot$ with a $2\sigma$
range of $ 4\times 10^{9} < M_\star < 1.9\times 10^{10}{\rm
  M}_\odot$. The age and star formation rate are obviously degenerate
with the star formation history and dust reddening, so no strong
constraints could be placed on those parameters. As was found in the
analysis of $\beta$, the UV spectral slope is fairly red and this
requires significant dust reddening. The best fit has dust extinction
of $A_{V}=0.75$ and the $2\sigma$ range is $0.48<A_{V}<1.48$.

\section{Galaxy luminosity function}
\label{lumfun}

\subsection{Completeness}
\label{completeness}

There are a number of factors that affect the completeness of the
sample, or equivalently, the effective volume of the survey. Firstly,
the sample does not contain every object brighter than a certain $z'$
magnitude limit due to incompleteness in the source detection
algorithm close to the limit. In addition, blending with brighter
objects becomes an issue in ground-based data at these faint
magnitudes. As was shown in the stacks of Section \ref{stacked}, there
is evidence for incompleteness of the sample due to foreground
contamination at $i'$ band which prevents some true high-redshift
galaxies from having $i'-z'>2$ in their aperture magnitudes.

\begin{figure}
\hspace{-0.25cm}
\resizebox{0.51\textwidth}{!}{\includegraphics{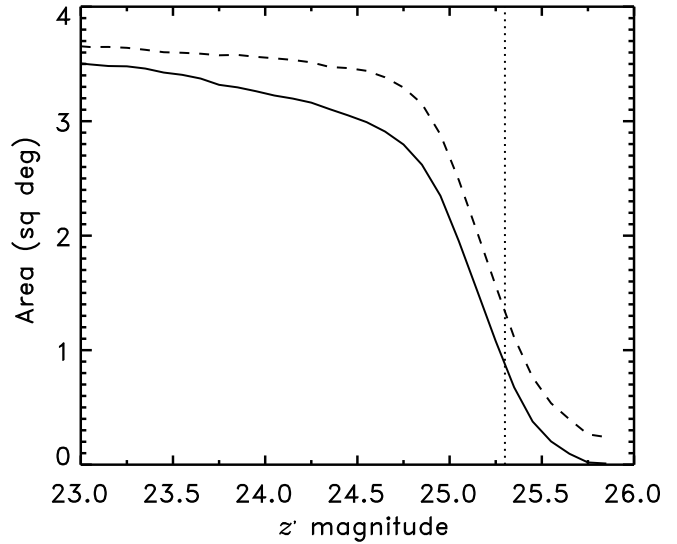}}
\caption{Effective survey area of the LBG sample as a function of magnitude
(solid line). Note that these magnitudes are the true magnitudes of
the simulated galaxies before photometric noise is added, so some
galaxies fainter than the magnitude limit ($z'=25.3$; dotted line) are
included in the sample. The dashed line shows what the effective area
would be without incompleteness due to foreground contamination of the
$i'$ band magnitude and hence color.\label{fig:zcomplete}
}
\end{figure}

In order to determine the effective sky area surveyed, we have carried
out an analysis of insertion and recovery of simulated sources into
the data. As found in Section \ref{sizes}, about 50\% of the $z=6$
LBGs are spatially resolved, whereas 50\% are consistent with point
sources. The simulated sources have a similar distribution of observed
FWHM to the LBGs. The selection criteria for recovery are identical to
those of the automated candidate selection. In addition, we check
whether there is a contaminating object at $i'$ band which would cause
the observed $i'-z'$ color to fall below the threshold. Figure
\ref{fig:zcomplete} shows the effective sky area surveyed in all four
Deep fields as a function of the input magnitude of the artificial
galaxies. The dashed line shows the effective area if one does not
consider the foreground contamination of the $i'-z'$ color. The
effective area declines slowly from $z'=23$ to $z'=24.8$ and then
falls more rapidly to $z'=25.5$.  Note that the effective area is
still considerable for input magnitudes $25.3<z'<25.5$, so some
galaxies fainter than the nominal magnitude limit will be in the
sample due to photometric scatter.

The next factors to consider for completeness are the effects of the
$i'-z'>2$ color criterion and the conversion from $z'$ magnitude to
absolute magnitude. These are assessed by considering the range of
properties expected for high-redshift galaxies, including
\lya\ emission line strength, UV continuum spectral slope and IGM
absorption of the UV continuum. As in Section \ref{spectra} we
use the $z=3$ LBG composite of Shapley et al. (2003) as the basis for
the galaxy template.  Absorption shortward of \lya\ due to foreground
neutral hydrogen is modeled as in Willott et al. (2010) using the mean
and scatter derived by Songaila (2004). 

A range of UV spectral slopes is implemented, as required by
observations. At $z\approx 6$ and high luminosity the reported typical
value in the literature is $\beta=-2$ (Bouwens et al. 2012;
Finkelstein et al. 2012). Our stacking and individual LBG fitting
results showed that for even more luminous $z\approx 6$ LBGs, the
typical $\beta=-1.4$. Combining our results with other work, we assume
a typical value for the population of $\beta=-1.8$ and a Gaussian
distribution with $1\sigma$ scatter of $0.5$. We have checked and
using a mean of $\beta=-1.5$ or $\beta=-2$ makes no difference to our results.

\begin{figure}
\resizebox{0.5\textwidth}{!}{\includegraphics{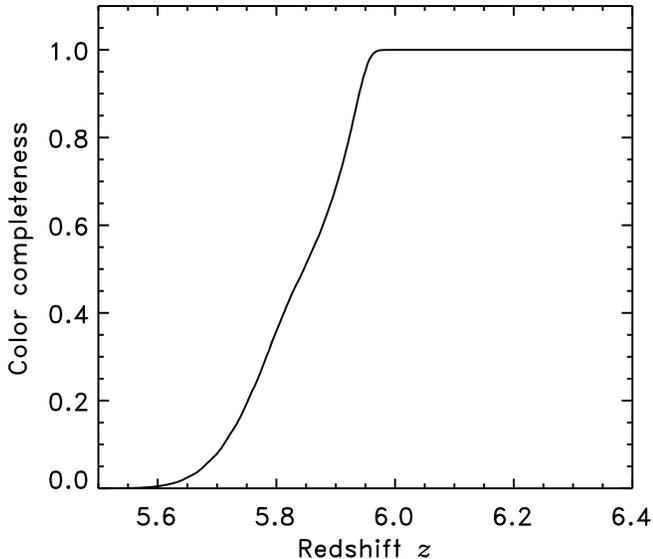}}
\caption{Completeness due to the color criterion of $i'-z'>2$ as a
  function of redshift. This curve was determined for a population of
  galaxies with \lya\ equivalent width, UV continuum slope, IGM
  absorption and photometric errors as described in the text. Because
  all objects with $i'-z'>2$ have been identified as either galaxies
  or brown dwarfs there is no high redshift cutoff in
  completeness corresponding to high $z'-J$ colors.\label{fig:colcomplete} }
\end{figure}

Several spectroscopic studies have investigated the distribution of
\lya\ equivalent widths at high redshifts. We combine the works of
Stark et al. (2011), Curtis-Lake et al. (2012) and our own work in
Section \ref{spectra} to derive a \lya\ rest-frame equivalent width
probability distribution for luminous $z\approx 6$ LBGs of $P(W_{{\rm
    Ly}\alpha}) \propto \exp(-W_{{\rm Ly}\alpha}/25)$. This is similar
to the exponential parameterization of Dijkstra \& Wyithe (2012), also
based on the data of Stark et al.

A sample of 500 simulated galaxies were generated with properties
randomly drawn from these distributions and their $i'z'J$ colors and
absolute magnitudes, $M_{1350}$, determined if located at all
redshifts between $z=5.5$ and $z=6.7$. Photometric scatter in the
colors is also included. The median $i'-z'$ and $z'-J$ color as a
function of redshift is plotted in Figure \ref{fig:izjcol}. This curve
runs very close to the stacked values of $i'-z'$ and $z'-J$,
suggesting the models are indeed a good representation of the data. At
redshifts $5.6\leq z \leq 6$, some fraction of galaxies are too blue
in $i'-z'$ to be included in the sample. In particular, we find that
galaxies with strong \lya\ lines at $z<5.9$ are much less likely to be
included in our sample than those with weak \lya\ lines. This could be
the reason why we find a relatively low \lya\ fraction for the
galaxies with best-fit redshifts $z \approx 5.8$ (c.f. Curtis-Lake et
al. 2011). At $z>5.9$, where this bias does not exist, the fraction
with strong \lya\ in our sample is quite high (two out of
three). Figure \ref{fig:colcomplete} shows the completeness as a
function of redshift due to the color selection criterion.

\subsection{Magnitude distributions}
\label{magdistrib}

\begin{figure}
\vspace{-0.3cm}
\resizebox{0.5\textwidth}{!}{\includegraphics{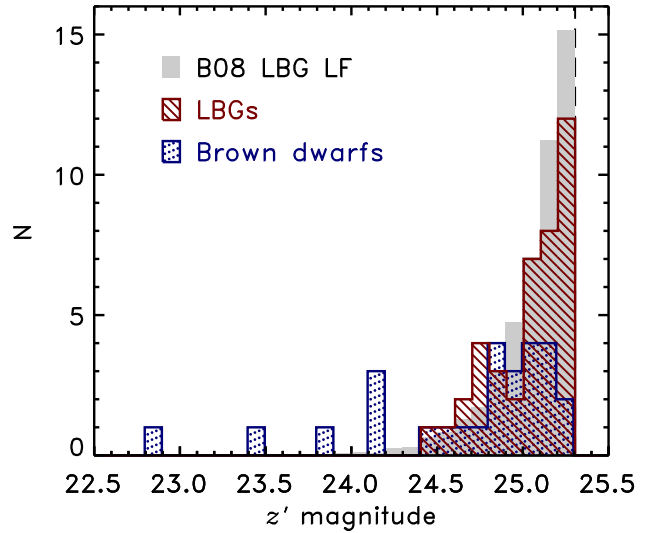}}
\caption{Magnitude histograms for Lyman break galaxy candidates and brown dwarfs selected by colors $i'-z'>2$ and $z'\leq 25.3$. The brown dwarf magnitude distribution (blue, dotted) has a fairly shallow slope due to the fact that the distance to L dwarfs is well beyond the disk scale length. The galaxy magnitude distribution (red, diagonal line) has a very steep slope. The gray shaded histogram shows the expected LBG counts for our survey based on the known luminosity function which is well-constrained at fainter magnitudes (Bouwens et al. 2008). The dashed black line shows the magnitude limit of our samples.
\label{fig:maghist}
}
\end{figure}

Before using these data to derive the $z=6$ galaxy luminosity
function, we consider simply the observed $z'$ band magnitude
distributions of the galaxies and brown dwarfs. Figure
\ref{fig:maghist} shows these two distributions. The dwarf magnitudes
extend to $z'=22.8$ and have a fairly flat distribution with a peak at
$z\approx 25.0$. This flat distribution is a consequence of galactic
structure and the survey completeness as a function of magnitude. At
magnitude $z'=25$, mid-L and mid-T dwarfs are 500 and 200 pc away,
respectively. The CFHTLS fields are at high galactic latitude and
hence the space density of brown dwarfs declines beyond the disk. The
galaxy distribution is quite different and much steeper. There are no
galaxies brighter than $z'=24.4$ and almost 70\% of the galaxies lie
in the small magnitude range of $25<z'<25.3$. The steepness of this
distribution, despite the declining survey completeness at faint
magnitudes, is due to the very steep bright end of the galaxy
luminosity function. 

The gray histogram in Figure \ref{fig:maghist} shows the expected
magnitude distribution for $z=6$ LBGs in our survey based on the
evolving luminosity function in Section 5.3 of Bouwens et
al. (2008). At $3.5 < z < 6.5$ this luminosity function is constrained
by the LBG luminosity functions determined in Bouwens et al (2007)
based on ACS observations in the GOODS, HUDF and HUDF-Parallel
fields. The expected counts take into account the selection criteria,
completeness and photometric uncertainty of our sample. It is clear
from Figure \ref{fig:maghist} that both the shape and overall
normalization of the expected counts are close to those observed in
our sample. This is remarkable given that our survey volume is 40
times greater than that used by Bouwens et al. (2007) and hence their
data contained very few sources as bright as ours. The expected
magnitude distribution predicts 46 LBGs in our sample, compared with
40 observed. The number expected at $z'<25$ is 13, identical to that
observed.

\subsection{Luminosity function derivation}

Because our LBG sample covers only a limited range of apparent and
absolute magnitudes, we cannot use it to determine the full galaxy
luminosity function at $z=6$. The luminosity function at the break and
at fainter magnitudes has already been well-studied from deep \hst
surveys over small sky areas (Bouwens et al. 2007). The main
contribution of our study is to determine the space density of very
rare, highly luminous LBGs.

Due to the fact that most CFHTLS LBGs have unknown redshifts
(within the Lyman break redshift selection range) and unknown
$k$-corrections due to \lya\ emission line contribution and spectral
slope, we cannot map a particular apparent magnitude onto an absolute
magnitude and redshift. Therefore we cannot carry out standard
luminosity function derivation methods where each observed source
corresponds to a point on the redshift, luminosity plane or by
counting the number of galaxies in luminosity bins. Instead we use the
stepwise maximum-likelihood method of Efstathiou et al. (1988) where
the luminosity function is characterized by values in a number of
absolute magnitude bins. We choose to use 5 regularly spaced nodes at
$M_{1350}=[-22.5,-22,-21.5,-21,-20.5]$ since this covers the full
extent of the absolute magnitude range of our detected LBGs and
ensures that extrapolation of the luminosity function beyond this
range will not bias our results.

\begin{deluxetable}{c c c c}
\hspace{-1.0cm}
\tablewidth{240pt}
\tablecolumns{4}
\tablecaption{\label{tab:lf} Luminosity function of $z\approx 6$ LBGs from CFHTLS} 
\tablehead{$M_{1350}$   & $\Phi$                   & $\Phi_{\rm low}$           & $\Phi_{\rm high}$\\
                       & (Mpc$^{-3}$\,mag$^{-1}$)  &   (Mpc$^{-3}$\,mag$^{-1}$) &   (Mpc$^{-3}$\,mag$^{-1}$)}
\startdata
$-22.5  $&$ 2.66\times 10^{-8}  $&$ 9.08\times 10^{-9}  $&$ 7.78\times 10^{-8}$\\ 
$-22.0  $&$ 2.18\times 10^{-6}  $&$ 8.70\times 10^{-7}  $&$ 9.70\times 10^{-6}$\\ 
$-21.5  $&$ 1.45\times 10^{-5}  $&$ 2.88\times 10^{-6}  $&$ 2.92\times 10^{-5}$\\ 
$-21.0  $&$ 1.29\times 10^{-4}  $&$ 7.06\times 10^{-5}  $&$ 2.19\times 10^{-4}$\\ 
$-20.5  $&$ 2.30\times 10^{-4}  $&$ 9.34\times 10^{-5}  $&$ 5.77\times 10^{-4}$ 
\enddata
\tablenotetext{a}{$\Phi_{\rm low}$ and $\Phi_{\rm high}$ are the lower and upper bounds for each absolute magnitude node containing 68\% of the bootstrap results.}
\end{deluxetable}

The comparison of the model with the data is performed in observed
magnitude space because this is the only way of accounting for the
selection effects and completeness discussed in
Section \ref{completeness} and the non-unique mapping of apparent to
absolute magnitude. For a model luminosity function (defined as
power-laws connecting and extending beyond the five nodes), the
completeness information and photometric noise are used to generate a
model observed $z'$ magnitude distribution, using the same method as
in Section \ref{magdistrib}. These model magnitudes are binned in
$\delta z'=0.1$ magnitude bins, $n_{\rm mod}$, and compared to the
observed $z'$ magnitude histogram, $n_{\rm obs}$, (Figure
\ref{fig:maghist}). The data are Poissonian and the likelihood,
$\mathcal{L}$, of observing the data given the model over all $N$
magnitude bins is given by
\begin{equation}
\mathcal{L}=\prod_{i=1}^{N} \frac{e^{-n_{{\rm mod}, i}} n_{{\rm mod}, i}^{n_{{\rm obs}, i}}}{n_{{\rm obs}, i}!}.
\end{equation}
The galaxy space densities at the five nodes are the five free
parameters which are adjusted using an amoeba algorithm (Press et
al. 1992) to determine the maximum likelihood. The 68\% range on the
best fit values are determined via bootstrapping by 400 trials of Poisson
perturbation of the $n_{{\rm obs}, i}$ values.

\begin{figure}
\hspace{-0.4cm}
\resizebox{0.52\textwidth}{!}{\includegraphics{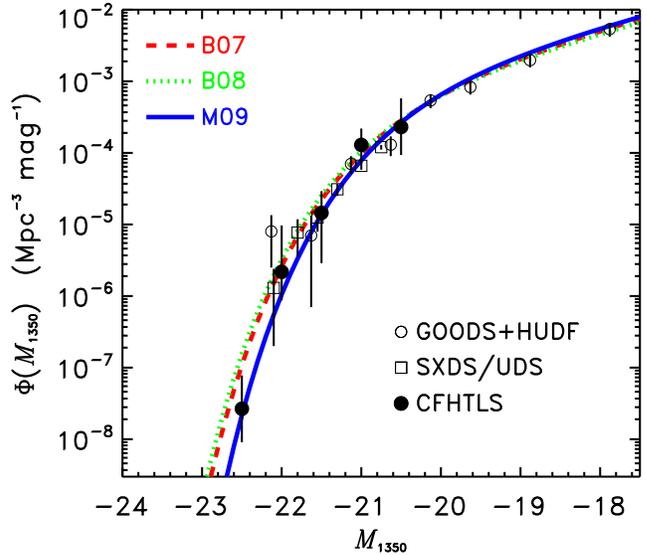}}
\caption{Galaxy luminosity function at $z\approx 6$. The maximum
  likelihood node points determined in this paper from CFHTLS are
  shown by filled black circles, whilst binned points from the
  literature are shown with open black symbols (GOODS+HUDF from
  Bouwens et al. 2007 and SXDS/UDS from McLure et al. 2009). Schechter
  function curves are plotted from the works of Bouwens et al. (2007)
  (red), Bouwens et al. (2008) (green) and McLure et al. (2009)
  (blue). The CFHTLS data at the bright end most closely match the
  McLure et al. derivation.
\label{fig:lf}
}
\end{figure}

The results are given in Table \ref{tab:lf} . Figure \ref{fig:lf} plots the data
and compares with previous results from the literature. Our best-fit
values match well the Schechter parameterizations of Bouwens et
al. (2007, 2008) and McLure et al. (2009) and agree with previous
binned values determined by Bouwens et al. (2007) and McLure et
al. (2009). Our two most luminous data points show the steep decline
in space density towards high luminosity required by our data. The
68\% uncertainties on our data points are fairly large due to the
small sample size and the fact that sources can be redistributed to
neighboring nodes in different trials. The total volume of our survey
is $\approx 10^7$\,Mpc. Therefore we are not truly measuring a space
density of $<10^{-7}$\,Mpc$^{-1}$ at $M_{1350}=-22.5$.  The low value
of this node is showing that there must be a sharp decline brighter
than $M_{1350}=-22$ or else we would observe more bright LBGs. One of
our spectroscopically confirmed galaxies, WMH\,5, has $M_{1350}=-22.65$, but
most of the rest have absolute magnitudes close to $M_{1350}=-22$, in
agreement with this sharp decline in number density.

\subsection{Field-to-field variance}

Part of the robustness of our results comes from the fact the CFHTLS
Deep is spread over four separate 1 square degree fields. We now
consider the variance in $z\approx 6$ LBG counts across the four
fields. We identified 8, 15, 8 and 9 LBGs in D1, D2, D3 and D4,
respectively. The two fields with the least certain identifications
due to shallower $J$-band (D3 and D4) do not have an unusual number of
counts, providing confidence in the LBG classification in these
fields. The field with the largest deviation is D2 (COSMOS) which,
with 15 LBGs, is 1.6$\sigma$ beyond the mean, assuming Poisson
variance only. Another source of variance is the large-scale
distribution of matter, so called cosmic variance. This could be
important if the most luminous LBGs are hosted by rare, massive dark
matter halos. We use the cosmic variance calculator of Trenti \&
Stiavelli (2008) assuming a duty cycle of 0.5 for the halos. For this
space density and duty cycle the minimum dark matter halo mass hosting
a galaxy is $2 \times 10^{12 }$\,M$_\odot$.  The variance calculation
shows that the cosmic variance contribution to the total variance in
each field is expected to be only half the Poisson variance
contribution. Hence including cosmic variance the D2 field is
1.4$\sigma$ beyond the mean and not unexpectedly overdense.

As well as chance, another factor that could be contributing to the
higher counts in D2 is that the CFHTLS data is a little shallower in
that field. With shallower data, there are three possible effects that
alter the expected number of candidates found: (i) the larger
magnitude uncertainties cause more objects to scatter in from
faintward of the $z'=25.3$ magnitude limit; (ii) the larger magnitude
uncertainties cause more objects to scatter in from blueward of the
$i'-z'=2$ color limit; (iii) the detection completeness will be lower
at the faint end. All of these effects are included for the sample as
a whole in Section \ref{completeness}, but we do not analyze them on a
field-by-field basis. Certainly, effect (i) is not significant because
comparison of the CFHTLS $z'$ and Subaru $z'$ magnitudes in D2 shows
that only two of the 15 have Subaru $z'>25.3$. A similar fraction is
found for D1 using VIDEO $Z$ magnitudes and is expected based on
Figure \ref{fig:zcomplete}. We do not have similar comparison data to
determine the likely strength of effects (ii) and (iii) (which affect
the number of candidates in opposite directions). Therefore we ascribe
the higher number of LBGs in D2 being mostly due to chance and the
total of 40 in the sample is unbiased and fully accounted for by our
selection function.

\subsection{Discussion}

Our results show that there is a sharp decline in the $z\approx 6$
galaxy space density for galaxies brighter than $M_{1350}=-22$,
consistent with the exponential function of Bouwens et al. (2007;
2008). The steepness of this function is much steeper than that of the
dark matter halo mass function for the expected host halo masses,
${M_H}$, of these luminous LBGs ($10^{12}<{M_H}<10^{13}\,{\rm M}_\odot$; Lee et
al. 2009; Trenti et al. 2010). This implies suppression of star
formation in the most massive halos at high redshift. The galaxy
luminosity and dark matter halo mass functions can be related using
the conditional luminosity function (CLF) method (Yang et
al. 2003). At low redshifts, the bright end of the galaxy luminosity
function is consistent with the relationship $L \sim M_H^{0.28}$ (Vale
\& Ostriker 2006). The physical explanation for this is likely
feedback from an AGN and/or inefficient gas cooling in high mass halos
(Benson et al. 2003).

Because our best-fit luminosity function is so similar to that of
Bouwens et al. (2007), we can adopt the $z=6$ CLF derivation of Trenti
et al. (2010) which was fit to this function. Trenti et al. (2010)
showed that by requiring only recently formed (200 Myr) halos to host
LBGs, the duty cycle for high mass halos must be almost unity, unlike
the values of $\approx 20$\% previously estimated for typical LBGs at
this epoch (Stark et al. 2007; Lee et al. 2009). Under this assumption
that the duty cycle does not vary much over the luminosity range of
interest, Trenti et al. (2010) showed that the high luminosity end of
the CLF follows a relationship of $L \sim M_H^{0.5}$. This is somewhat
steeper than is found at low redshift, but still indicates that
luminosity is a slowly varying function of halo mass.

Models with star formation efficiency and duty cycle (or star
formation timescale) which are independent of halo mass do not predict
such a steep decline in the luminosity function at the brightest
magnitudes (Stark et al. 2007; Mu\~noz \& Loeb 2011). These works were
able to fit the $z=6$ luminosity function of Bouwens et al. (2007)
because those data did not provide evidence of the sharp exponential
decline. Our results show that the same process that limits star
formation in high mass halos at low redshift is also acting at a time
just one billion years after the Big Bang.

\section{Conclusions}
\label{conc}

We have performed the largest area survey for luminous LBGs at
$z\approx 6$. These galaxies have been used to study their physical
sizes and spectral energy distributions. The large sky area has
enabled the most robust results on the space density decline at the
bright end of the luminosity function. Below we list the specific
conclusions drawn from this work.

\begin{itemize}

\item Spectroscopy of seven galaxies revealed redshifts for four of
  them. Three of these have \lya\ emission lines, one of which is
  extremely weak with $W_{{\rm Ly}\alpha} = 4$\AA. All four galaxies
  show continuum redward of \lya\ which enables redshift determination
  from the break location.

\item About half of the LBGs are resolved in the ground-based CFHT
  imaging observations. This is consistent with a typical intrinsic
  half-light radius for luminous $z\approx 6$ galaxies of 2\,kpc,
  higher than the typical 1\,kpc in GOODS. High resolution imaging of
  a subsample shows that some galaxies have a dominant component with
  large half-light radii and some have multiple components. A larger
  sample with high resolution imaging is necessary to determine if
  there is a relationship between interaction frequency and
  star-formation rate.

\item The properties of the stacked photometry have been investigated.
  These data show a flux decrement in the observed-frame optical of
  nearly 100 due to the Lyman break. The $YJHK$ photometry are fit by
  a power-law of slope $\beta=-1.44 \pm 0.10$, much redder than for less
  luminous galaxies at this epoch.  The stacked SED is well fit by a
  $z=5.95$ constant star formation model with mass $\approx
  10^{10} {\rm M}_\odot$. Significant dust reddening is required
  ($A_V=0.75$) showing that dust is more prevalent in the most massive
  systems, again possibly due to these being merger-induced
  starbursts.

\item The magnitude distributions of our LBG sample and a $i'-z'$
  color-matched sample of L/T dwarfs are considered. Whilst the dwarfs
  have a fairly flat distribution with a peak at $z'\approx 25$, the
  LBGs have a distribution that rises sharply with magnitude. Our
  $z\approx 6$ luminosity function derivation provides clear evidence
  for a sharp decline in space density brighter than $M=-22$.

\item The decline in space density at high luminosity is consistent
  with a relationship between galaxy luminosity and dark matter halo
  mass of $L \sim M_H^{0.5}$. This shows that the processes limiting
  star formation in high mass halos at low redshift are also operating
  effectively at a time just 1 billion years after the Big Bang.

\end{itemize}

\acknowledgments

This paper is dedicated to the memory of Steve Rawlings, who taught
CJW to look for the signal in the noise. Thanks to Genevieve Soucail
for providing reduced near-IR data for some of our targets. Thanks to
the anonymous referee for suggestions that considerably improved the
paper. Based on observations obtained with MegaPrime/MegaCam, a joint
project of CFHT and CEA/DAPNIA, at the Canada-France-Hawaii Telescope
(CFHT) which is operated by the National Research Council (NRC) of
Canada, the Institut National des Sciences de l'Univers of the Centre
National de la Recherche Scientifique (CNRS) of France, and the
University of Hawaii. This work is based in part on data products
produced at TERAPIX and the Canadian Astronomy Data Centre as part of
the Canada-France-Hawaii Telescope Legacy Survey, a collaborative
project of NRC and CNRS. We gratefully acknowledge use of data from
the ESO Public Survey programs 179.A-2005 and 179.A-2006 with the
VISTA telescope.This work uses observations taken by the CANDELS
Multi-Cycle Treasury Program with the NASA/ESA HST, which is operated
by the Association of Universities for Research in Astronomy, Inc.,
under NASA contract NAS5-26555. Based on observations obtained at the
Gemini Observatory, which is operated by the Association of
Universities for Research in Astronomy, Inc., under a cooperative
agreement with the NSF on behalf of the Gemini partnership: the
National Science Foundation (United States), the Particle Physics and
Astronomy Research Council (United Kingdom), the National Research
Council (Canada), CONICYT (Chile), the Australian Research Council
(Australia), CNPq (Brazil) and CONICET (Argentina).  This paper uses
data from Gemini programs GN-2009A-Q-2, GS-2009A-Q-3, GN-2010B-C-9 and
GN-2011A-C-1.


\begin{thebibliography}{}

\bibitem[]{} Allard, F., \& Freytag, B. 2010, Highlights of Astronomy, 15, 756
\bibitem[]{} Barkana, R. \& Loeb, A. 2001, Phys. Rep. 349, 125
\bibitem[]{} Barmby, P., Huang, J.-S., Ashby, M. L. N., et al. 2008, ApJS, 177, 431
\bibitem[]{} Benson, A., Bower, R. G., Frenk, C. S., Lacey, C. G., Baugh, C. M., \& Cole, S. 2003, ApJ, 599, 38
\bibitem[]{} Bertin, E., \& Arnouts, S. 1996, A\&AS, 117, 393
\bibitem[]{} Bielby, R., Hudelot, P., McCracken, H. J., et al. 2012, A\&A, 545A, 23
\bibitem[]{} Bouwens, R. J., Illingworth, G. D., Blakeslee, J. P., Broadhurst, T. J., \& Franx, M. 2004, ApJ, 611, L1
\bibitem[]{} Bouwens, R. J., Illingworth, G. D., Blakeslee, J. P., \& Franx, M. 2006, ApJ, 653, 53
\bibitem[]{} Bouwens, R. J., Illingworth, G. D., Franx, M., \& Ford, H. 2007, ApJ, 670, 928
\bibitem[]{} Bouwens, R. J., Illingworth, G. D., Franx, M., \& Ford, H. 2008, ApJ, 686, 230
\bibitem[]{} Bouwens, R. J., Illingworth, G. D., Oesch, P. A., et al. 2012, ApJ, 754, 83
\bibitem[]{} Bruzual, G., \& Charlot, S. 2003, MNRAS, 344, 1000
\bibitem[]{} Bunker, A. J., Stanway, E. R., Ellis, R. S., \& McMahon, R. G. 2004, MNRAS, 355, 374
\bibitem[]{} Burgasser, A. J., Reid, I. N., Siegler, N., et al. 2007, in Protostars and Planets V, ed. B. Reipurth, D. Jewitt, \& K. Keil (Tucson, AZ: Univ. Arizona Press), 427
\bibitem[]{} Calzetti D., Armus L., Bohlin R., Kinney A., Koornneef J., \& Storchi- Bergmann T. 2000, ApJ, 533, 682
\bibitem[]{} Capak, P., Mobasher, B., Scoville, N. Z., et al. 2011, ApJ, 730, 68
\bibitem[]{} Cole, S. 1991, ApJ, 367, 45
\bibitem[]{} Curtis-Lake, E., McLure R. J., Pearce, H., et al. 2012, MNRAS, 422, 1425
\bibitem[]{} Delorme, P., Willott, C. J., Forveille, T., et al. 2008, A\&A, 484, 469 
\bibitem[]{} Dickinson, M., Stern, D., Giavalisco, M., et al. 2004, ApJ, 600, L99
\bibitem[]{} Dijkstra, M., \& Wyithe, J. S. B. 2012, MNRAS, 419, 3181
\bibitem[]{} Dow-Hygelund, C. C., Holden, B. P., Bouwens, R. J., et al. 2007, ApJ, 660, 47
\bibitem[]{} Dunlop, J. S., McLure, R. J., Robertson, B. E., et al. 2012, MNRAS, 420, 901
\bibitem[]{} Dutton, A. A., van den Bosch, F. C., Faber, S. M., et al. 2011, MNRAS, 410, 1660
\bibitem[]{} Efstathiou, G., Ellis, R. S., \& Peterson, B. A. 1988, MNRAS, 232, 431
\bibitem[]{} Ferguson, H. C., Dickinson, M., Giavalisco, M., et al. 2004, ApJ, 600, L107
\bibitem[]{} Finkelstein, S. L., Papovich, C., Salmon, B.,  et al. 2012, ApJ, 756, 164
\bibitem[]{} Finlator, K., Oppenheimer, B. D., \& Dav\'e, R. 2011, MNRAS, 410, 1703
\bibitem[]{} Grogin, N. A., Kocevski, D. D., Faber, S. M., et al. 2011, ApJS, 197, 35
\bibitem[]{} Hayes, M., Laporte, N., Pell\'o, R., Schaerer, D., \& Le Borgne, J.-F. 2012, MNRAS, 425, L19 
\bibitem[]{} Jarosik, N., Bennett, C. L., Dunkley, J., et al. 2011, ApJS, 192, 14
\bibitem[]{} Jarvis, M. J., Bonfield, D. G., Bruce, V. A., et al. 2012, MNRAS, submitted, arXiv:1206.4263
\bibitem[]{} Jiang, L., Egami, E., Kashikawa, N., et al. 2011, ApJ, 743, 65 
\bibitem[]{} Kennicutt, R. C. 1998, ARA\&A, 36, 189
\bibitem[]{} Koekemoer, A. M., Faber, S. M., Ferguson, H. C., et al. 2011, ApJS, 197, 36
\bibitem[]{} Lee, K.-S., Giavalisco, M., Conroy, C., Wechsler, R. H., Ferguson, H. C., Somerville, R. S., Dickinson, M. E., \& Urry, C. M. 2009, ApJ, 695, 368
\bibitem[]{} Lonsdale, C. J., Smith, H. E., Rowan-Robinson, M., et al. 2003, PASP, 115, 897
\bibitem[]{} McCracken, H. J., Milvang-Jensen, B., Dunlop, J. S., et al. 2012, A\&A, 544A, 156 
\bibitem[]{} McLure R. J., Jarvis M. J., Targett T. A., Dunlop J. S., \& Best P. N., 2006, MNRAS, 368, 1395
\bibitem[]{} McLure R. J., Cirasuolo M., Dunlop J. S., Foucaud S., \& Almaini O., 2009, MNRAS, 395, 2196
\bibitem[]{} Miller, S. H., Bundy, K., Sullivan, M., Ellis, R. S., \& Treu, T. 2011, ApJ, 741, 115
\bibitem[]{} Mu\~noz, J. A., \& Loeb, A.  2011, ApJ, 729, 99
\bibitem[]{} Nagao, T., Murayama, T., Maiolino, R., et al. 2007, A\&A, 468, 877
\bibitem[]{} Peng, C. Y., Ho, L. C., Impey, C. D., \& Rix, H.-W. et al. 2010, AJ, 139, 2097
\bibitem[]{} Press, W. H., Teukolsky, S. A., Vetterling, W. T., \& Flannery, B. P. 1992, Numerical Recipes in C: The Art of Scientific Computing (Cambridge: Cambridge Univ. Press)
\bibitem[]{} Reyl\'e C., Delorme, P., Willott, C. J., et al. 2010, A\&A, 522, A112
\bibitem[]{} Sanders, D. B., Salvato, M., Aussel, H., et al. 2007, ApJS, 172, 86
\bibitem[]{} Scoville, N., Abraham, R. G., Aussel, H., et al. 2007, ApJS, 172, 38
\bibitem[]{} Shapley, A. E., Steidel, C. C., Pettini, M., \& Adelberger, K. L. 2003, ApJ, 588, 65
\bibitem[]{} Shimasaku, K., Ouchi, M., Furusawa, H., et al. 2005, PASJ, 57, 447
\bibitem[]{} Songaila, A. 2004, AJ, 127, 2598
\bibitem[]{} Spinrad, H., Stern, D., Bunker, A., Dey, A., Lanzetta, K., Yahil, A., Pascarelle, S., \& Fern\'andez-Soto, A. 1998, AJ, 116, 2617
\bibitem[]{} Stark, D. P., Loeb, A., \& Ellis, R. S., \& Ouchi, M. 2007, ApJ, 668, 627
\bibitem[]{} Stark, D. P., Ellis, R. S., \& Ouchi, M. 2011, ApJ, 728, L2
\bibitem[]{} Su, J., Stiavelli, M., Oesch, P., et al. 2011, ApJ, 738, 123
\bibitem[]{} Taniguchi, Y., Scoville, N., Murayama, T., et al. 2007, ApJS, 172, 9
\bibitem[]{} Trenti, M., \& Stiavelli, M. 2008, ApJ, 676, 767 
\bibitem[]{} Trenti, M., Stiavelli, M., Bouwens, R. J., Oesch, P., Shull, J. M., Illingworth, G. D., Bradley, L. D., \& Carollo C. M. 2010, ApJ, 714, L202
\bibitem[]{} Vale, A., \& Ostriker, J. P. 2006, MNRAS, 371, 1173
\bibitem[]{} Vanzella, E., Giavalisco, M., Dickinson, M., et al. 2009, ApJ, 695, 1163
\bibitem[]{} White, R. L., Helfand, D. J., Becker, R. H., Glikman, E., \& de Vries, W. 2007, ApJ, 654, 99
\bibitem[]{} Wilkins, S. M., Bunker, A. J., Stanway, E., Lorenzoni, S., \& Caruana, J. 2011, MNRAS, 417, 717
\bibitem[]{} Willott, C. J., Delorme, P., Reyl\'e, C., et al. 2009, AJ, 137, 3541                 
\bibitem[]{} Willott, C. J., Delorme, P., Reyl\'e, C., et al. 2010, AJ, 139, 906  
\bibitem[]{} Yan, H., \& Windhorst, R. A. 2004, ApJ, 612, L93
\bibitem[]{} Yang, X., Mo, H. J., \& van den Bosch, F. C. 2003, MNRAS, 339, 1057
\end{thebibliography}
\end{document}